\documentclass[a4paper,12pt]{article}
\pdfoutput=1 

\usepackage{jheppub} 

\usepackage[T1]{fontenc} 

\usepackage{float, array, xspace, amscd, amsthm}
\usepackage{comment}
\usepackage{mathrsfs}
\usepackage{fancyhdr}
\usepackage{longtable}

\newcommand{\dd}{{\rm d}}
\newcommand{\w}{\wedge}
\newtheorem{theorem}{Theorem}
\newtheorem{lemma}{Lemma}
\newtheorem{proposition}{Proposition}

\newcommand{\IC}{\mathbb{C}}

\newcommand{\IH}{\mathbb{H}}

\newcommand{\IO}{\mathbb{O}}

\newcommand{\IR}{\mathbb{R}}

\newcommand{\IZ}{\mathbb{Z}}

\newcommand{\scC}{\mathscr{C}}
\newcommand{\scS}{\mathscr{S}}
\newcommand{\scT}{\mathscr{T}}

\newcommand{\cG}{\mathcal{G}}
\newcommand{\cL}{\mathcal{L}}

\newcommand{\cT}{\mathcal{T}}
\newcommand{\cV}{\mathcal{V}}
\newcommand{\be}{\begin{equation}}
\newcommand{\ee}{\end{equation}}
\def\bea#1\eea{\begin{align}#1\end{align}}
\newcommand{\tr}{{\rm tr}}

\DeclareMathOperator{\Ima}{Im}
\DeclareMathOperator{\Rea}{Re}
\newcommand{\Maps}{{\rm Maps}}
\usepackage{color}
\usepackage[dvipsnames]{xcolor}

\title{
Almost contact structures on manifolds with a $G_2$ structure}
\author[a]{Xenia de la Ossa,}
\author[b]{Magdalena Larfors,}
\author[b]{Matthew~Magill.}

\affiliation[a]{Mathematical Institute, Oxford University\\Andrew Wiles Building, Woodstock Road\\Oxford OX2 6GG, UK }
\affiliation[b]{Department of Physics and Astronomy,Uppsala University\\ SE-751 20 Uppsala, Sweden}

\emailAdd{delaossa@maths.ox.ac.uk, magdalena.larfors@physics.uu.se, matthew.magill@physics.uu.se}

\abstract{ We review the construction of almost contact metric (three-) structures, abbreviated ACM(3)S, on manifolds with a $G_2$ structure. These are of interest for certain supersymmetric configurations in string and M-theory. We compute the torsion of the $SU(3)$ structure associated to an ACMS and apply these computations to heterotic $G_2$ systems and supersymmetry enhancement.  We initiate the study of the space of ACM3Ss, which is an infinite dimensional space with a local product structure and interesting topological features.  Tantalising links between ACM3Ss and associative and coassociative submanifolds are observed.  }

\begin{document}

\maketitle
\flushbottom

\newpage

\section{Introduction}
Low-dimensional, minimally supersymmetric vacua of string theory and M-theory are of interest for a number of reasons. In four dimensions, such ground states may provide effective theories for the physical world, and give mathematically consistent UV completions of the standard models of particle physics and cosmology. Similarly, the AdS/CFT correspondence provides further motivation to explore supersymmetric vacua in spacetimes with constance negative curvature in both three and four dimensions.   More broadly, both four- and three-dimensional solutions can serve as toy models where dualities, deformations and geometric invariants may be explored through the lens of superstring theory. 

The class of manifolds with $G_2$ structure, which includes torsion-free $G_2$ manifolds as a subclass, holds an important place in this field of research. In M-theory they provide  four-dimensional $N=1$ Minkowski vacua, and in type II or heterotic string theory they give rise to three-dimensional vacua with non-positive cosmological constant. These topics have been studied for a number of years, and important results have been established. In particular, large classes of torsion-free $G_2$ manifolds have been constructed \cite{joyce1996:1,joyce1996:2,kovalev20,Corti:2012kd,Corti:2013} and studied in detail by mathematicians and physicists. However, the topic of $G_2$ structures and related superstring compactifications is far from exhausted.  For example, one still lacks a deformation theory of different $G_2$ geometries, and $G_2$ instantons, beyond the infinitesimal level established in Refs.~\cite{joyce1996:1,joyce1996:2,Gutowski:2001fm,Grigorian:2009ge} and \cite{delaOssa:2016ivz,delaOssa:2017pqy,Fiset:2017auc,delaOssa:2018azc,Clarke:2016qtg,Clarke:2020erl}; there are no direct constructions\footnote{However, see e.g. \cite{Braun:2017uku} for constructions relying on string duality.} of the singular, compact $G_2$ manifolds that are needed for the existence of chiral families in M-theory vacua \cite{Atiyah:2001qf,Acharya:2001gy}; a method for counting associative submanifolds, important in physics applications due to their contributions to the non-perturbative superpotential, remains to be established \cite{Joyce:2016fij,braun2018infinitely,Acharya:2018nbo,Harvey:1999as}; and proofs are lacking for conjectures regarding duality and  mirror symmetry among $G_2$ vacua 
\cite{gukov2003duality,Braun:2017uku,braun2018towards,braun2017mirror,Eckhard:2018raj}.

In this paper we discuss the existence of almost contact (three-) structures on $G_{2}$ structure manifolds. In brief, an almost contact structure (ACS) is related to a nowhere vanishing vector field, and an almost contact three-structure (AC3S) is related to a triple of such vector fields (the formal definition is given below). Both structures are guaranteed to exist on any $G_2$ structure manifold \cite{thomas1969}. Our aim with this paper is, in part, to provide a detailed review of these well-established mathematical facts, which, to our knowledge, has been been partly lacking (see however \cite{friedrich1997nearly} and \cite{Behrndt:2005im}). This may explain why the ACS perspective has not been emphasized in the recent physics and mathematics literature related to $G_2$ structure manifolds.  In addition, we hope our paper gives evidence that these almost contact (three-) structures provide useful perspectives on $G_2$ geometry, calibrated submanifolds, string vacua and supersymmetry.

Almost contact structures are, in some way, odd-dimensional cousins of almost complex structures in even dimensions. Just as for almost complex structures in even dimensions, the almost contact structures give rise to projection operators and a decomposition of e.g. differential forms into longitudinal and transverse components. More precisely, almost contact structures induce almost complex structures on the transverse geometry.

As mentioned above, ACS are related to nowhere-vanishing vector fields. The existence of nowhere vanishing vector fields, or, more generally, nowhere vanishing differential forms, on a manifold indicates that the tangent bundle structure group can be reduced. In string compactifications, this is intimately tied to the amount of supersymmetry preserved by the vacuum. One may thus suspect that the existence of an AC(3)S may lead to string solutions with extended supersymmetry and we will demonstrate under which conditions this holds true. More generally, almost contact structures provide additional information that may be used in the classification of supersymmetric string vacua. Indeed, while their related AC(3)S has not necessarily been emphasised, $SU(2)$ and $SU(3)$ structures have been used in classifications of four-dimensional $N=1$ M-theory vacua \cite{Behrndt:2005im}, {in constructing M-theory lifts of $N=1$ type IIA solutions \cite{Andriolo:2018yrz}},  $N=1$ AdS type IIB vacua  \cite{Kim:2005ez,Gran:2007ps,Passias:2019rga}, and has points in common with the Gran--Papadopoulos classification of heterotic supersymmetric vacua \cite{Gran:2005wf,Gran:2007kh,Gran:2016zxk}.

Nowhere-vanishing vector fields are always allowed in odd dimensions; it is well known that the obstruction to the existence of a nowhere vanishing vector field on a closed manifolds is a non-vanishing Euler characteristic  \cite{Hopf27}.\footnote{In this paper, we will focus in particular on closed (i.e. compact, boundaryless) manifolds. The results we state also hold for non-compact manifolds, and manifolds with boundary.} %
Odd-dimensional manifolds have vanishing Euler characteristic, and hence admit an almost contact structure. It is less obvious, but nonetheless true, that seven-dimensional manifolds admit three linearly independent, nowhere vanishing vector fields \cite{10.2307/45277146,thomas1969,kuo1970}. Moreover, as we will explain below, when combined with the positive three-form that defines a $G_2$ structure, these vector fields give rise to an almost contact three-structure, which is furthermore compatible with the metric induced by a $G_2$ structure \cite{friedrich1997nearly,Arikan:2012acs,Arikan:2011acs,Todd:2015era}. Importantly, the AC3S vectors will not, in general, be parallel with respect to the $G_2$ connection. Thus, while this shows that $G_2$ structure manifolds are necessarily reduced to $SU(2)$, there is no automatic reduction of the holonomy of the $G_2$ connection. A further reduction in the holonomy would be indicative of enhanced supersymmetry and, therefore, not generically expected.

In the rest of this paper we will provide a more detailed account of almost contact metric three-structures on manifolds with $G_2$ structures. {This is in part a review of established facts, in part a derivation of new results.} We start in section \ref{sec:prelnot} by briefly reviewing the background material and setting our notation. In section \ref{sec:ACMS} we {review} almost contact metric structures and the associated $SU(3)$ structure using {a somewhat novel} perspective of differential forms. We also {add new observations on this topic by elaborating} on the foliation associated to the vector field of the ACMS, and the related transverse six dimensional geometry, which indeed carries an $SU(3)$ structure with intrinsic torsion that we may directly determine. With this result we can, in section  \ref{sec:hetACMS}, show how $N=1$ heterotic $G_2$ systems may be analysed using ACMS. We then turn, in section \ref{sec:ACM3S}, to { a review of} almost contact metric three-structures and their associated $SU(2)$ structure. {Building on these classical results, we expand, in section \ref{sec:count-ac3s}  upon} the existence of three- and four-dimensional foliations related to ACM3Ss and discuss the intriguing relation between such foliations and associative and coassociative submanifolds (which  are also calibrated submanifolds if the $G_2$ structure is closed and coclosed{, respectively}). {This allows us to} {initiate a study of} the space of ACM3Ss and show that it has a non-trivial structure. In section \ref{sec:examples} we give a number of examples which illustrate the concepts we have reviewed. Finally, in section \ref{sec:conc} we conclude and point out a number of directions for future studies.

\subsection{Preliminary notions} \label{sec:prelnot}
\subsubsection{$G_2$ structures}\label{ssec:g2struct}

A $G_2$ structure manifold is a seven dimensional manifold $Y$ along with a reduction of the tangent bundle structure group to $G_2$ (see for example \cite{FerGray82, Bryant:2005mz} and \cite{joyce2000} for more details on $G_2$ structures). A $G_2$ structure exists whenever the seven manifold is orientable and spin. 

Since the group $G_2$ can be identified with the group of automorphisms of the imaginary octonions, reducing the structure group to $G_2$ allows us to identify the tangent bundle as a bundle of imaginary octonions, ${\rm Im}\,\IO$. Making such an identification endows $Y$ with a metric and a vector cross product, which can in turn be encoded in a stable, positive three-form, $\varphi$.

The relation between these data is as follows. Given a metric, $g$ and cross-product, $\times$, the three-form is defined by
\begin{equation} 
    \varphi(u,v,w)=g(u\times v,w)\,.\label{eq:g2crossprod}
\end{equation}
Locally, one can choose a trivialisation of the tangent bundle in which such a three-form takes a standard form, which in our conventions will be
 \be \label{eq:g2standard}
\varphi_0 = (e^{12}+e^{34}+e^{56})\w e^7 + e^{135}-e^{146}-e^{236}-e^{245}\,.
\ee 

On the other hand, whenever a three-form can be put in such a form, we can extract a metric and cross product. Indeed, once we have a metric,  \eqref{eq:g2crossprod} defines a cross product. The metric can be defined, using a choice of orientation, by
\begin{equation}
6 g_\varphi(u, v)\, \dd {\rm vol}_\varphi
= (u\lrcorner\varphi)\wedge(
v\lrcorner\varphi)\wedge\varphi~,
\label{eq:g2metric}
\end{equation}
for all vectors $u$ and $v$ in $\Gamma(TY)$. 
In components this means
\[
g_{\varphi\, ab} = \frac{\sqrt{\det g_\varphi}}{3!\, 4!}\, 
\varphi_{a c_1 c_2}\, \varphi_{b c_3 c_4}\, \varphi_{c_5 c_6 c_7}\,
\epsilon^{c_1\cdots c_7}
=  \frac{1}{4!}\, 
\varphi_{a c_1 c_2}\, \varphi_{b c_3 c_4}\, 
\psi^{c_1 c_2 c_3 c_4}~,\]
where we have used the metric to define a dual four-form
\[ \psi = *_\varphi\,\varphi~,\]
and 
\[ \dd x^{a_1\cdots a_7} = \sqrt{\det g_\varphi}
\ \epsilon^{a_1\cdots a_7}\, \dd {\rm vol}_\varphi~.\]
With respect to this metric, the three-form $\varphi$, and hence its Hodge dual $\psi$, are  normalised so that
\[ \varphi\wedge *\varphi = ||\varphi||^2\, \dd{\rm vol}_\varphi
~, \qquad ||\varphi||^2= 7~,\]
that is
\[ \varphi\lrcorner\varphi = \psi\lrcorner\psi = 7
~.
\]

Choosing a local frame where $\varphi$ takes its standard form, one can verify that the metric, $g_\varphi$, is the standard Euclidean metric and the four-form will be
\begin{equation}\label{eq:psistandard}
    \psi_0 = e^{3456} + e^{1256}+ e^{1234} - e^{2467}+e^{2357}+e^{1457}+e^{1367}
\end{equation}

The exterior derivative of the forms $(\varphi,\psi)$, which give the structure equations for the $G_2$ structure, can be decomposed into irreducible representations of $G_2$  
\begin{align}
\dd_7\varphi &= \tau_0 \, \psi + 3\, \tau_1\wedge\varphi + *\tau_3~,\label{eq:g2phi}
\\
\dd_7\psi&= 4\, \tau_1\wedge\psi  - \tau_{2}\wedge \varphi
~,\label{eq:g2psi}
\end{align}
where the torsion classes $\tau_k$ are $k$ forms,
$\tau_3$ is in the $\bf 27$ irreducible representation of $G_2$ and $\tau_{2}$ in the $\bf 14$.

\subsubsection{Almost contact structures}\label{ssec:ACS}

Let $Y$ be an odd dimensional Riemannian manifold with metric $g$.  If the manifold $Y$  admits the existence of an endomorphism $J$ of the tangent bundle $TY$, a unit vector field $R$ (with respect to the metric $g$), and a one-form $\sigma$ which satisfy
\[
J^2 = - {\bf 1} + R\otimes\sigma~, \quad\sigma(R) = 1~,
\]
$Y$ is said to admit an {\it almost contact structure} $(J, R, \sigma)$ \cite{sasaki1960, sasaki1961}. The one-form $\sigma$ is called the {\it contact form}. The dimension of a manifold admitting an ACS must be odd and the structure group of the tangent space reduces to $U(n)\times {\bf 1} $, where $2n+1$ is the dimension of $Y$.  
The ACS on the manifold $Y$ is said to be a  {\it contact structure} if 
\[
\sigma\wedge\dd\sigma\wedge\cdots\dd\sigma \ne 0~,
\]
everywhere on $Y$. 
In this paper we are mostly interested in the existence of almost contact structures on manifols with a $G_2$ structure.

A Riemannian manifold  $Y$  with an ACS $(J, R, \sigma)$ has an  {\it almost contact metric structure}  $(J, R, \sigma, g)$ (ACMS) if moreover  
\be
g(Ju, Jv)= g(u,v) - \sigma(u)\, \sigma(v)~, \qquad\forall u, v \in \Gamma(TY)~,
\ee
is satisfied.  The {\it fundamental two-form} $\omega$ of an almost contact metric manifold is defined by
\be
\omega(u,v) = g(Ju, v)~, \qquad\forall\, u, v \in \Gamma(TY)~,
\ee
and satisfies
\be
\sigma\wedge\omega\wedge\cdots\wedge\omega \ne 0~.
\ee

An {\it almost contact 3-structure} (AC3S) on a manifold $Y$ \cite{kuo1970} is defined by three 
distinct almost contact structures $(J^\alpha, R^\alpha, \sigma^\alpha),~\alpha = 1, 2, 3$ on $Y$ which satisfy the following conditions
\be \label{eq:ac3sintro}
\begin{split}
J^\gamma &= J^\alpha\, J^\beta - R^\alpha\otimes \sigma^\beta = - J^\beta J^\alpha + R^\beta\otimes \sigma^\alpha~,
\\
R^\gamma &= J^\alpha(R^\beta) = - J^\beta(R^\alpha)~,
\\
\sigma^\gamma &= \sigma^\alpha \circ  J^\beta = - \sigma^\beta \circ J^\alpha~,
\\
\quad \sigma^\alpha(R^\beta) &= \sigma^\beta(R^\alpha) = 0~,
    \end{split}
\ee
where $\{\alpha, \beta,\gamma\}$ are a cyclic permutation of $\{1, 2, 3\}$.  A manifold admitting an AC3S must have dimension $4n+3$ where $n$ is a non-negative integer and the structure group of the tangent space reduces to $Sp(n)\times {\bf 1}_3$. An {\it almost contact metric 3-structure} on a Riemannian manifold $Y$ with metric $g$, is an AC3S which satifies
\be
g(J^\alpha u, J^\alpha v) = g(u,v) - \sigma^\alpha(u)\, \sigma^\alpha(v)~, \forall~ u, v \in \Gamma(TY)~,
\ee
for each $\alpha\in\{1,2,3\}$. An AC3S consisting of three contact structures satisfying \eqref{eq:ac3sintro} is a contact 3-structure and defines a 3-Sasakian geometry \cite{kashiwada}. 

\section{$SU(3)$ structures on manifolds with a $G_2$ structure}\label{sec:ACMS}

A manifold $Y$ with a $G_2$ structure $\varphi$ has more structure than expected.  In this section we review the fact
that  $(Y,\varphi)$ admits an almost contact metric structure (ACMS) and thereby reduces the structure group to $SU(3)$ \cite{Arikan:2012acs,Arikan:2011acs,Todd:2015era}.\footnote{In fact, we will see in section \ref{sec:ACM3S} that there are least three non-zero vectors on a manifold with a $G_2$ structure \cite{thomas1969}, which gives  $Y$ an almost contact metric 3-structure (ACM3S) inducing an $SU(2)$ structure on $Y$ \cite{friedrich1997nearly}. } The reason this happens stems from the fact that any such manifold admits a nowhere vanishing vector field \cite{Hopf27} $R$ which can be normalized with respect to the $G_2$ metric $g_\varphi$.  

In the first two subsections we prove the following proposition.

\begin{proposition}\label{prop:restrict}
\cite{friedrich1997nearly,Todd:2015era} Let $(Y,\varphi)$ be a seven dimensional manifold $Y$ with a $G_2$ structure $\varphi$ and let $g_{\varphi}$ 
be the metric on $Y$ determined by $\varphi$.  Then $Y$ admits an ACMS $(J, R, \sigma, g_\varphi)$ determined  by a unit vector field $R$, where the endomorphism $J$ is given by
\[
J(u)= R\times_\varphi u~, \quad\forall~u\in \Gamma(TY)~,
\]
and the one form $\sigma$ dual to $R$ with respect to the $G_2$ metric $g_\varphi$
\[ \sigma(R) = 1~.\]
The ACMS determines a foliation ${\cal F}_R$ of $Y$ by the one dimensional integral curves of $R$ with $G_2$ metric
\[
\dd s_\varphi^2 = \sigma^2 + \dd s_\perp^2~,
\]
where $\dd s_\perp^2$ represents the metric on the transverse geometry of ${\cal F}_R$  induced by the ACMS on $Y$. 

Furthermore, the ACMS induces a reduction of the $G_{2}$ structure to an $SU(3)$ structure $(\omega_\varphi,\Omega)$ on 
the transverse geometry of the foliation, where $\omega_\varphi$ is the fundamental two form on $Y$ 
\[
\omega_\varphi = i_R(\varphi)~,
\]
and $\Omega$ is a transverse three form of type $(3,0)$ with respect to $J$.  Both forms $(\omega_\varphi, \Omega)$
are determined uniquely by the ACS decomposition of the $G_2$ structure $\varphi$ on $Y$
\be
\varphi = \sigma\wedge\omega_\varphi + \Omega_+~.\label{eq:phiACS}
\ee
The coassociative four form $\psi$ dual to $\varphi$ decomposes as
\[
\psi  = *_\varphi\varphi =  - \sigma\wedge\, \Omega_- + \frac{1}{2}\,\omega_\varphi\wedge\omega_\varphi
~.
\]
\vskip10pt

\end{proposition}

Notice that there is no guarantee that an ACMS will be compatible with a $G_2$ connection, i.e. if $\nabla$ is a given connection on $Y$ with ${\rm Hol}(\nabla)\subseteq G_2$, it does not follow from a choice of ACMS that ${\rm Hol}(\nabla)\subseteq SU(3)$. These issues are discussed in section \ref{ssec:decomstructeq}.  In appendix \ref{app:su3struct} we review the definition of manifolds with an $SU(3)$ structure.

\subsection{Almost contact metric structures on a manifold with a $G_2$ structure}\label{ssec:g2ACS}

Let $Y$ be a seven dimensional compact manifold.  It is known that there exists (at least) one nowhere vanishing vector field, $R$, on $Y$ \cite{Hopf27}.  This  vector field defines a one dimensional ``characteristic foliation'' ${\cal F}_R$ of $Y$, where the one dimensional leaves are the integral curves of $R$. 
Given the non-vanishing vector field $R$ on $Y$, we can choose local coordinates on $Y$ adapted to the foliation structure
\be
\{x^a, a = 1,\ldots 7\} = \{(r, x^m), m = 1\ldots 6\}~,\label{eq:coords}
\ee 
 such that $r$ is the coordinate along the integral curves of $R$ (curves of constant $x^m$) and such that the vector $R$ is given by
\be
R =  \partial_r~.\label{eq:R}
\ee

Suppose now that $Y$ has a $G_2$ structure $\varphi$ with metric $g_\varphi$. Without loss of generality, we choose the vector $R$ to be of unit length with respect to the $G_2$ metric. We define a unique one form $\sigma$ by
\be
\sigma(u) = g_\varphi(R, u)~, \qquad\forall\, u\in \Gamma(TY)
~.\label{eq:sigma}
\ee
Note that 
\be
\sigma(R) = 1~,\label{eq:unitR}
\ee
is the statement that $R$ has unit length in terms of $\sigma$. 

The $G_2$ structure together with a choice of vector field, $R$, defines an endomorphism, $J$, of $TY$ by
\be
 J (u) = R \times_\varphi u~, \qquad \forall\ u\in \Gamma(TY)
~, \label{eq:JY}
\ee
where $\times_\varphi$ is the cross product on $Y$ determined by the $G_2$ structure
\be
\varphi(u, v, w) = g_\varphi(u \times_\varphi v, w)~,
\qquad \forall\ u, v, w \in \Gamma(TY)~.
\label{eq:cross}
\ee
Equivalently, we can consider $J$ to be a vector-valued one form on $Y$, in which case it is given, in local coordinates, by
\be
J^a{}_b = - \varphi^a{}_{bc}\, R^c = - i_R(\varphi)^a{}_b~.\label{eq:Jcomps}
\ee
Then, one can easily prove that 
\be 
J^2 = - {\bf 1} + R\otimes \sigma~,\label{eq:J2}
\ee
using the cross product identity
\be
u \times_\varphi (u \times_\varphi v) = - g_\varphi(u,u)\, v + g_\varphi(u,v)\, u~,
\qquad \forall\ u, v \in \Gamma(TY)~. \label{eq:idcross}
\ee
Moreover,
\be
J(R)= 0~, \quad {\rm and}\qquad \sigma(J(u)) = 0~, \quad \forall\ u\in \Gamma(TY)~.\label{eq:ortho}
\ee
Therefore \cite{Todd:2015era} the $G_2$ structure on $Y$,  together with the existence of a (unit length) nowhere vanishing vector field $R$ on $Y$ determine an {\it almost contact structure} $(J, R, \sigma)$ on $Y$ (see section \ref{ssec:ACS}).  The one form $\sigma$ is the contact form.   The existence of the almost contact structure (ACS) on $Y$ means that
there is a reduction of the structure group $G_2$  to ${\bf 1} \times U(3)$.  We will see this explicitly in this section.
We remark that this ACS on $Y$ is not, in general, a contact structure as it need not be that $\sigma$ satisfies everywhere on $Y$ the condition
\[
\sigma\wedge\dd_7\sigma\wedge\dd_7\sigma\wedge\dd_7\sigma\ne 0~.
\]

Furthermore \cite{Arikan:2012acs,Arikan:2011acs,Todd:2015era}, 
this ACS is compatible with the metric, in the sense that that $g_\varphi$ satisfies the necessary condition
\be
g_\varphi(Ju, Jv) = g_\varphi(u,v) - \sigma(u)\, \sigma(v)~,
\qquad \forall~ u, v \in \Gamma(TY)~.\label{eq:acms}
\ee
  and therefore $(J, R, \sigma, g_\varphi)$ defines an ACMS. 

Given the almost contact metric structure $(J, R, \sigma, g_\varphi)$ on $Y$, one also has the fundamental two form $\omega_\varphi$ on $Y$
\be
\omega_\varphi(u, v) = g_\varphi (Ju, v) = \varphi(R, u, v)~, \qquad \forall\ u, v \in \Gamma(TY)~.\label{eq:fundform}
\ee
Equivalently, 
\be
\omega_\varphi = i_R(\varphi)~.\label{eq:iRphi}
\ee
This two form indeed satisfies
\[ 
\sigma\wedge\omega_\varphi\wedge \omega_\varphi\wedge \omega_\varphi \ne 0~,
\]
because
\be
 \dd{\rm vol}_{\varphi} = \frac{1}{6}\, \sigma\wedge\omega_\varphi\wedge \omega_\varphi\wedge \omega_\varphi~,
 \label{eq:u3vol}
\ee
which can be easily proven from \eqref{eq:g2metric} and \eqref{eq:iRphi}.

Given the ACS $(J, R, \sigma)$ on $Y$,  consider  the bundle, ${\rm Ker}(\sigma)$, whose fibres are the vectors which are orthogonal to $R$.  
This is a codimension one subbundle of $TY$ and induces an orthogonal decomposition of $TY$ as
\[
TY = {\rm Span}\{R\} \oplus {\rm Ker}(\sigma)~, \qquad
{\rm Ker}(\sigma)= {\rm Span}\{R\}^\perp
~.
\]
 The endomorphism $J$ maps sections of  $\Gamma(TY)$ into sections of ${\rm Ker}(\sigma)$, as can be seen from equation \eqref{eq:ortho},  and thus it can be used to construct projection operators on $TY$.  
Indeed, the operator
\[
  -\, J^2 = {\bf 1} - R\otimes\sigma~,
\]
is a projection operator mapping  $TY$ into 
${\rm Ker}(\sigma)$.  The action of the projection operators can naturally be extended to the cotangent bundle $T^*Y$, which now has the orthogonal decomposition
\[
T^*Y = {\rm Span}\{\sigma\} \oplus {\rm Span}\{\sigma\}^\perp~,
\]
and, therefore, to any tensor on $Y$. We can uniquely decompose any $k$-form $\alpha$ as
\be
\alpha = \sigma\wedge\alpha_{0} + \alpha_{\perp}~, \label{eq:kform}
\ee
where $\alpha_{0}$ and $\alpha_{\perp}$ are respectively a $(k-1)$-form and a $k$-form
on $Y$ such that
\[
i_{R}(\alpha)= \alpha_{0}~, \qquad i_{R}(\alpha_{0})= 0~,
\qquad i_{R}(\alpha_{\perp})=0~.
\]

Furthermore, the endomorphism $J$ induces an orthogonal decomposition of  ${\rm Ker}(\sigma)$ over $\IC$ 
\[
{\rm Ker}(\sigma) \otimes \IC 
= {\rm Ker}(\sigma)_{\IC}{}^{(1,0)} \oplus {\rm Ker}(\sigma)_{\IC}{}^{(0,1)}~.
\]
In fact, there are also projection operators
$P$ and $Q$ on ${\rm Ker}(\sigma)\otimes \IC$  
\begin{align*}
P &= \frac{1}{2}\, ( {\bf 1} - i \, J - R\otimes\sigma)
= - \frac{i}{2}\, J\, ({\bf 1} - i J)~,
\\[5pt]
Q &
= \frac{1}{2}\, ( {\bf 1} + i \, J - R\otimes\sigma)
= + \frac{i}{2}\, J\, ({\bf 1} + i \, J)~,
\end{align*}
which map  ${\rm Ker}(\sigma)$ into  ${\rm Ker}(\sigma)_{\IC}{}^{(1,0)}$ or  ${\rm Ker}(\sigma)_{\IC}{}^{(0,1)}$ respectively. 

Again, the action of these operators can naturally be extended to  $T^*Y$
and to  any tensor on $Y$.  
For the  cotangent bundle $T^*Y$, ${\rm Span}\{\sigma\}^\perp\otimes\IC$  has the orthogonal decomposition
\[
{\rm Span}\{\sigma\}^\perp\otimes\IC = 
\left({\rm Span}\{\sigma\}^\perp\right)^{(1,0)}
\oplus\left({\rm Span}\{\sigma\}^\perp\right)^{(0,1)}~.
\]
We can decompose any $k$-form $\alpha$ with
respect to the ACS as in equation \eqref{eq:kform} where the forms $\alpha_{0}$
and $\alpha_{\perp}$  decompose further into $(p,q)$-type with respect to $J$.

We have arrived at the conclusion that the $G_{2}$ structure on the manifold $Y$ is reduced to ${\bf 1}\times U(3)$.  The $U(3)$
structure on $Y$ is determined by the endomorphism $J$ which is effectively an almost complex structure on $Y$.  Moreover, we have a fundamental  form $\omega_{\varphi}$
satisfying \eqref{eq:u3vol}.   We discuss below in more detail how this works and how the $U(3)$ structure reduces further to an $SU(3)$ structure. %

\subsection{Transverse geometry and $SU(3)$ structures on $Y$}
\label{sec:ACStrans}

In this section we show explicitly how the structure group $G_2$ is reduced to $SU(3)$.  The fact that any $G_2$ structure manifold admits a reduction to an $SU(3)$ structure group was already shown by \cite{friedrich1997nearly}. In that paper, the authors  argued that a nowhere vanishing vector field  $R$ on $(Y,\varphi)$, along with the nowhere vanishing spinor $\eta$ implicit in the choice of $G_2$ structure, induce a second spinor by Clifford multiplication, $R\eta$. These spinors can be used to construct the $SU(3)$ structure on $Y$.  In this section we instead show how the structure group is reduced to $SU(3)$ by constructing the $SU(3)$ structure directly from the $G_{2}$ structure three form $\varphi$ and the ACS.\footnote{Note that, although there is  always a connection $\nabla$ with $G_2$ holonomy on $Y$  such that $\nabla\eta = 0$, it is not necessarily the case that  
$\nabla(R\eta) = 0$, and therefore the holonomy group of $\nabla$ is not necessarily reduced to $SU(3)$.}  A summary of the content of this section and the previous one is given in proposition \ref{prop:restrict} at the beginning of this section \ref{sec:ACMS}.

We begin by reviewing the notion of the {\it transverse geometry} of the foliation ${\cal F}_R$. 
Loosely speaking, locally the transverse geometry pretends to be the geometry of a hyperplane $X$ which is transverse to $R$. Note, however, that there is not necessarily a six dimensional manifold $X\subset Y$ whose tangent plane is transverse to $R$. Another way to say this is that ${\rm Ker}(\sigma)$ is not necessarily integrable.

We say that a vector $u\in \Gamma(TY)$ is {\it transverse} to the foliation ${\cal F}_R$ if $u\in \Gamma({\rm Ker}(\sigma))$. As we saw above, the vector $J(u)$ is transverse, for any $u\in TY$ (see equation \eqref{eq:ortho}). For a $k$-form given as in equation \eqref{eq:kform}, we can think of $\alpha_0$ and $\alpha_\perp$ as a  $k-1$ form and a $k$  form respectively on the transverse geometry of the foliation on $Y$.  
We will call $\alpha_\perp$ the {\it transverse} component of $\alpha$, and we call a form {\it transverse}  if
\be
i_R(\alpha) = 0~.\label{eq:transvform}
\ee

Clearly, $J(\alpha)$ is transverse for any $k$-form $\alpha$ on $Y$ as
$i_R(J(\alpha)) = 0$. 
The endomorphism $J$ restricts to an almost complex structure, $J_\perp$ on the transverse geometry of ${\cal F}_R$, that is, $J_\perp$ is an endomorphism of ${\rm Ker}(\sigma)$ with $ J_\perp^2 = -{\bf 1}~$.  The action of $J$ on any $k$ form $\alpha$ on $Y$ can be written as 
\[
J(\alpha) = J(\alpha_\perp) = J_\perp(\alpha_\perp)
~.
\]
 The transverse geometry then carries all the properties of an almost complex structure.  For instance,  as mentioned before any transverse form decomposes into $(p,q)$-type with respect to $J_\perp$.
Furthermore, the fundamental form $\omega_\varphi$ is transverse, and one can define a hermitian structure on the transverse geometry.  It is not hard to prove that $\omega_\varphi$ is type $(1,1)$ with respect to $J_\perp$,
that is
\[
J(\omega_\varphi) = \omega_\varphi~.
\]
One can define {\it primitive} transverse forms on $Y$ and thus define the Lefshetz decomposition, with respect to $\omega_\varphi$, of transverse forms.

Consider the metric $g_\varphi$ on $Y$.  Using equation \eqref{eq:fundform} and the endomorphism $J$, one can establish a metric $g_\perp$ on the 
transverse geometry. In fact, \eqref{eq:fundform} and \eqref{eq:J2} imply that
on $Y$, the $G_2$ metric is given by
\be
g_\varphi(u,v) =  \omega_\varphi(u, Jv)+\sigma(u)\, \sigma(v)~,
\qquad u,\, v\, \in\, \Gamma(TY)~,
\ee
where the fact that $R$ is unit length is clearly satisfied.  Also, if $u\in \Gamma({\rm Ker}(\sigma))$, then we have by \eqref{eq:sigma} that
\[
g_\varphi(u,R)  = \sigma(u) = 0~.
\]
Now, let $u\, , v\, \in \Gamma({\rm Ker}(\sigma))$.  Then we {\it define} a metric $g_\perp$ on the transverse geometry by
\be
g_\varphi(u,v) =  \omega_\varphi(u, Jv) =  g_\perp(u,v)~.
\label{eq:gtransv}
\ee
This means we have a ``bundle like'' metric on $Y$ of the form
\be
\dd s_\varphi^2 = \sigma^2 + \dd s_\perp^2~,\label{eq:7metric}
\ee
where $\dd s_{\perp}^2$ represents the line element on the transverse geometry 
and in the coordinate system adapted to the foliation ${\cal F}_{R}$,
the one form $\sigma$  can be written as
\be
\sigma = \dd r + \Sigma~, \qquad \Sigma = \Sigma_{m}\, \dd x^{m}~.
\ee
The transverse one form $\Sigma$ behaves like a one form connection on the transverse geometry as can be verified by performing a coordinate transformation on $Y$.  Now, recall that the $G_2$ structure $\varphi$ determines a metric $g_\varphi$ uniquely on $Y$ by equation \eqref{eq:g2metric}.  
Of course, this metric needs to be the same as that in equation \eqref{eq:7metric}. 

To compare \eqref{eq:7metric} with \eqref{eq:g2metric}, we begin by decomposing $\varphi$ with respect to the ACMS $(J, R, \sigma, g_\varphi)$. Recall that the ACMS determines detemines the fundamental two form on $Y$
\[
\omega_\varphi = i_R(\varphi) 
~.
\]
This means that $\varphi$ can be  decomposed uniquely as
\be
\varphi = \sigma\wedge \omega_\varphi +  \Omega_+~.
\label{eq:phidecomp}
\ee
for some well defined transverse real three form $\Omega_+$ on $Y$. Let $u\, , v\,  \in \Gamma(TY)$ and  consider this decomposition of  $\varphi$ together with equation \eqref{eq:g2metric} for the metric:
\[
6\, g_\varphi(u,v) \, \dd{\rm vol}_\varphi = i_u(\varphi)\wedge i_v(\varphi)
\wedge \varphi~.
\]
Taking $u = v =R$, and using $g_\varphi(R, R) = 1$, we have
\[
6\,\dd{\rm vol}_\varphi = i_R(\varphi)\wedge i_R(\varphi)\wedge\varphi
= \omega_\varphi \wedge\omega_\varphi \wedge( \sigma\wedge\omega_\varphi  + \Omega_+)~.
\]
The last term must vanish as it is a transverse seven form, 
while the first term gives the volume form on $Y$ in terms of $\sigma$ and $\omega_\varphi$
\be
\dd{\rm vol}_\varphi = \frac{1}{6}\, \sigma\wedge\omega_\varphi \wedge\, \omega_\varphi \wedge\omega_\varphi ~.\label{eq:g2vol}
\ee
We have already seen this before, see \eqref{eq:u3vol}.
Consider now the metric components for $u\in \Gamma({\rm Ker}(\sigma))$ and $v = R$.  As in this case $g_\varphi(u, R)= \sigma(u)= 0$, we have
\begin{align*}
0 &= \sigma\wedge\big(  i_u(\Omega_+)
\wedge\omega_\varphi
- i_u(\omega_\varphi)\wedge\Omega_+\big)\wedge\omega_\varphi
\\
&= \sigma \wedge \left( 
i_{u}(\Omega_{+}\wedge\omega_\varphi\wedge\omega_\varphi)
+ 3\, i_{u}(\omega_\varphi)\wedge\Omega_{+}\wedge\omega_\varphi
\right)
= \sigma \wedge 3\, i_{u}(\omega_\varphi)\wedge\Omega_{+}\wedge\omega_\varphi
~,
\end{align*}
which must be true for all transverse vectors $u\in\Gamma({\rm Ker}(\sigma))$.  Hence
\be
\omega_\varphi\wedge\Omega_+ = 0~.\label{eq:su3one}
\ee
An important consequence of this equation \eqref{eq:su3one},  is that $\Omega_+$ is a primitive form type of $(3,0)+(0,3)$ with respect to $J$ because
$\omega_\varphi$ is type $(1,1)$.
Finally we calculate the components
$g_\varphi(u, v)$ where $u$ and $v$ are both in $\Gamma({\rm Ker}(\sigma))$. In this case,
$g_\varphi(u,v)= g_\perp(u,v)$ and we have
\[
\begin{split}
6\, g_\perp(u,v) \, \dd{\rm vol}_\varphi 
&= \sigma\wedge\Big(
i_u(\Omega_+)\wedge i_v(\Omega_+)\wedge\omega_\varphi
\\
&\qquad\qquad
- \big(i_u(\omega_\varphi)\wedge i_v(\Omega_+) + 
i_v(\omega_\varphi)\wedge i_u(\Omega_+)
\big)\wedge\Omega_+
\Big)
\\
&=
3\,\sigma\wedge\Omega_+\wedge
 \,i_u(\Omega_+)\wedge\, i_v(\omega_\varphi)
 ~,
\end{split}
\] 
where we have used the constraint \eqref{eq:su3one}.  Using the volume form \eqref{eq:g2vol} and contracting with $R$, we find
\be
\frac{1}{6}\, g_\perp(u,v) \, \omega_\varphi\wedge\omega_\varphi\wedge\omega_\varphi
= \frac{1}{2}\, \Omega_+\wedge
 \,i_u(\Omega_+)\wedge\, i_v(\omega_\varphi)~,
 \quad u\, , v\, \in \Gamma({\rm Ker}(\sigma))
 ~.\label{eq:vollambda}
\ee
As $i_u(\Omega_+)$ is a transverse two form type $(2,0) + (0,2)$, for any  vector $u$, it is easy to see that
\[
i_u(\Omega_+) = - J_\perp\big(i_u(\Omega_+)\big) = i_{J_\perp u}\big(J_\perp(\Omega_+)\big)~, \quad\forall\, u\in\Gamma({\rm Ker})(\sigma)
~.
\]
Then
\[
\begin{split}
\frac{1}{3}\, g_{\perp}(u,v) \, \omega_\varphi\wedge\omega_\varphi\wedge\omega_\varphi
&= \Omega_+\wedge
 \, i_{J_\perp u}\big(J_\perp(\Omega_+)\big)
\wedge\, i_v(\omega_\varphi)
\\
&=
- i_{J_\perp u} \Big(\Omega_+\wedge J_\perp(\Omega_+)
\wedge\, i_v(\omega_\varphi)\Big)
 + i_{J_\perp u}(\Omega_+)\wedge J_\perp(\Omega_+)
\wedge\, i_v(\omega_\varphi)
\\
&\qquad\qquad
+ \Omega_+\wedge J_\perp(\Omega_+)~ i_{J_\perp u}i_v(\omega_\varphi)
\\
&=  g_{\perp}(u, v)\, \Omega_+\wedge J_\perp(\Omega_+)
+ J_\perp(\Omega_+)\wedge i_{J_\perp u}(\Omega_+)
\wedge\, i_v(\omega_\varphi)~,
\end{split}
\]
where, by equation \eqref{eq:gtransv}
\[
i_{J_\perp u}i_v(\omega_\varphi) = - \omega_\varphi(Ju, v) = g_{\perp}(u,v)~.
\]
Consider now the last term.  Using the fact that $J_\perp^2 = - {\bf 1}$ and that
the action of $J_\perp$ on a transverse six form is the identity, we find
\[
\begin{split}
J_\perp(\Omega_+)\wedge i_{J_\perp u}(\Omega_+)\wedge \, i_v(\omega_\varphi) &=
- J_\perp\Big(
- \Omega_+\wedge\, J_\perp\big(i_{J_\perp u}(\Omega_+)\big) \wedge\, J_\perp\big(i_v(\omega_\varphi)\big)
\Big)
\\
&= 
- \Omega_+\wedge\, i_{J_\perp u}(\Omega_+) \wedge\, i_{J_\perp v}(\omega_\varphi)
\end{split}
\]
By equations \eqref{eq:vollambda} and \eqref{eq:acms}
\[
J_\perp(\Omega_+)\wedge i_{J_\perp u}(\Omega_+)\wedge \, i_v(\omega_\varphi)
= - \frac{1}{3}\, g_\varphi(Ju, Jv)\, \omega_\varphi\wedge\omega_\varphi\wedge\omega_\varphi
= - \frac{1}{3}\, g_\varphi(u, v)\, \omega_\varphi\wedge\omega_\varphi\wedge\omega_\varphi
~.
\]
Therefore
\[
\frac{1}{6}\, g_{\perp}(u, v)\, \omega_\varphi\wedge\omega_\varphi\wedge\omega_\varphi
= \frac{1}{4}\, g_{\perp}(u, v)\, \Omega_+\wedge J_\perp(\Omega_+)~,
\]
that is,
\be
\frac{1}{6}\, \omega_\varphi\wedge\omega_\varphi\wedge\omega_\varphi
= \frac{1}{4}\, \Omega_+\wedge J_\perp(\Omega_+)~.\label{eq:su3comp0}
\ee
Let $\Omega$ be a $(3,0)$ form with respect to $J_\perp$ such that
\be
\Omega_+ = \frac{1}{2}\, (\Omega + \overline\Omega) = {\rm Re}\,\Omega~.\label{eq:reOm}
\ee
Then we have that
\[
J_\perp(\Omega_+) = \frac{1}{2\, i}\, (\Omega - \overline\Omega) = {\rm Im}\,\Omega = \Omega_-~.
\]
Then equation \eqref{eq:su3comp0} becomes
\be
\frac{1}{6}\, \omega_\varphi\wedge\omega_\varphi\wedge\omega_\varphi
= \frac{1}{4}\, \Omega_+\wedge \Omega_- = \frac{i}{8}\, \Omega\wedge\overline\Omega~.\label{eq:su3comp}
\ee

One can define a Hodge-dual operator $*_\perp$ on the transverse geometry using the transverse metric $g_\perp$.  Let $\alpha$ be a $k$-form on $Y$ with
decomposition
\[
\alpha = \sigma\wedge\alpha_0 + \alpha_\perp~,
\]
then, it is not too difficult to show that 
\be
*_\varphi\, \alpha = (-1)^k\, \sigma\wedge\, *_\perp\, \alpha_\perp + *_\perp\, \alpha_0~,\label{eq:hodge*}
\ee
where one needs,
\[
\det{g_\varphi} = \sigma_0^2\, \det{g_\perp}~. 
\]
The coassociative four form $\psi$  then decomposes as
\be
\psi  = *_\varphi\varphi =  - \sigma\wedge\, \Omega_- + \rho_\varphi~,\label{eq:psiACS}
\ee
where we have defined the transverse four form $\rho_\varphi$ by
\be
\rho_\varphi = *_\perp\,\omega_\varphi = \frac{1}{2}\, \omega_\varphi\wedge\omega_\varphi~,\label{eq:rho}
\ee
and used the fact that
\[
*_\perp\Omega_+ = \Omega_- ~.
\]
This ends the proof of proposition \ref{prop:restrict}.

It is interesting to ask at this point about the conditions for the transverse geometry to correspond in fact to a submanifold $X\subset Y$. According to the Frobenius Theorem, the foliation ${\cal F}_R$ has a global transverse section $X$ when ${\rm Ker}(\sigma)$ is an integrable distribution, that is
\be
[u, v] \in \Gamma({\rm Ker}(\sigma))~, \quad\forall~ u\, , v\,  \in \Gamma({\rm Ker}(\sigma))~.\label{eq:frob}
\ee
A short computation shows that this is the case if and only if
\[
(\dd_7\sigma)(u,v) = 0~, \quad\forall~ u\, , v\,  \in \Gamma({\rm Ker}(\sigma))~,
\]
that is,
\be
\dd_7\sigma = \sigma\wedge\alpha~,\label{eq:dualfrob}
\ee
for some transverse one form $\alpha$, so $\dd_7\sigma$ has no transverse component. 
We will come back to this constraint later when we discuss examples (see section \ref{sec:nilman}).

\subsection{Decomposing the structure equations}\label{ssec:decomstructeq}

Let $Y$ be a manifold with a $G_{2}$ structure given by the three form $\varphi$, and let $g_{\varphi}$ be the metric on $Y$ determined by $\varphi$.  In this subsection,  we decompose the $G_{2}$ structure equations \eqref{eq:g2phi} and \eqref{eq:g2psi} under the ACMS $(J, R,  \sigma, g_{\varphi})$, to obtain the torsion classes $W_{i}$ of the induced transverse $SU(3)$ structure in terms of those $\tau_{i}$ of the $G_{2}$ structure.
We recall that under the ACMS $(J, R, \sigma, g_\varphi)$ the  $G_2$ structure is decomposed in terms of the underlying transverse $SU(3)$ structure $(\omega_\varphi, \Omega_+)$ as
\begin{align}
\varphi &= \sigma\wedge\omega + \Omega_+~,\\
\psi &= - \sigma\wedge\Omega_- + \rho~,\\
\sigma &= \dd r + \Sigma~,
\end{align}
where from now on we drop the label $\varphi$ in the fundamental two form.

In order to  decompose the structure equations under the ACMS, we begin with the following lemmas regarding the decomposition of the contraction operator $\lrcorner_\varphi$ and the exterior derivative.  The proof of both lemmas is straightforward.

\begin{lemma}\label{lemma:lrcorner}
Let $\alpha$ be a $p$-form and $\beta$ a $(p+q)$-form on $Y$. Let
\[
\alpha = \sigma\wedge\alpha_0 + \alpha_\perp~, \qquad \beta = \sigma\wedge\beta_0 + \beta_\perp~,
\]
be their decomposition with respect to the ACS. Then
\[
\alpha\lrcorner_\varphi\,\beta
=
(-1)^p\, \sigma\wedge \,(\alpha_\perp\,\lrcorner\,\beta_0)
+  (\alpha_0\lrcorner\beta_0
+ \alpha_\perp\,\lrcorner\,\beta_\perp)~,
\]
where $\lrcorner_\varphi$ and $\lrcorner$ are the contraction operators with respect to the $G_2$ metric $g_\varphi$ and the transverse metric $g_\perp$ respectively.

\end{lemma}

\begin{lemma}\label{lemma:derivs}
Let $\alpha$ be a transverse p-form on $Y$, that is, $i_R(\alpha) = 0$.  The decomposition under the ACMS of the exterior derivative $\dd_7\alpha$  is given by
\be
\dd_7\alpha = \sigma\wedge R(\alpha) + \dd_\perp\alpha~,\label{eq:d7ontransv}
\ee 
where we have defined
\be
\dd_\perp\alpha = \dd\alpha - \Sigma\wedge R(\alpha)~,\label{eq:dperp}
\ee
and $\dd$ refers to derivatives with respect to the transverse coordinates in the coordinate system $\{r, x^m\}$ adapted to the one dimensional foliation of $Y$ by the non-zero vector $R$ (see \eqref{eq:coords}).  The operator $\dd_\perp$ is a derivation, in particular it satisfies the Leibnitz rule, and its curvature $\dd^2_\perp$ can be understood as the curvature of $\dd_\perp$ given by
its action on transverse forms by
\be
\dd_\perp^2 = - \dd_\perp\Sigma\wedge R~.\label{eq:curvdperp}
\ee  
The exterior derivative $\dd_7$ on the one-form $\sigma$ is then
\be
\dd_7 \sigma = \dd_7 \Sigma = \sigma\wedge R(\Sigma) + \dd_\perp \Sigma~,
\label{eq:d7sigma}
\ee
Finally, equations \eqref{eq:d7ontransv}-\eqref{eq:d7sigma} imply that the exterior derivative $\dd_7\Lambda$ of a $p$-form $\Lambda$ on $Y$ with components
\[
\Lambda = \sigma\wedge\Lambda_0 + \Lambda_\perp~,
\]
is given by
\be
\dd_7\Lambda = \sigma\wedge \big( - \dd\Lambda_0  + R(\Lambda_\perp + \Sigma\wedge\Lambda_0)\big)
+ \dd_\perp \Lambda_\perp + \dd_\perp \Sigma\wedge\Lambda_0~.\label{eq:d7decomp}
\ee
\end{lemma}

Using Lemma \ref{lemma:derivs}, the decomposition of $\dd_7\varphi$ and $\dd_{7}\psi$ is given by
\begin{align}
\dd_7\varphi &= \sigma\wedge \big( -\dd\omega + R(\Omega_+ + \Sigma\wedge\omega) \big) + \dd_\perp \Omega_+ + \dd_\perp\Sigma\wedge \omega~,\label{eq:d7varphidecomp}
\\
\dd_{7}\psi &= \sigma\wedge \big( \dd\Omega_{-} + R(\rho - \Sigma\wedge\Omega_{-}) \big) + \dd_{\perp}\rho - \dd_{\perp}\wedge\Omega_{-}~.\label{eq:d7psidecomp}
\end{align}
Equating these with the decomposition under the ACMS of the right hand side of the $G_{2}$ structure equations  \eqref{eq:g2phi} and \eqref{eq:g2psi},  we find the following relations for the transverse $SU(3)$ structure
\begin{align*}
\dd\Omega_{+} &= \tau_{0}\, \rho  + 3 \, \tau_{1\, \perp}\wedge \Omega_{+} - \big(J(\tau_{3\, 0})+ \dd\Sigma\big)\wedge\omega
+ \Sigma\wedge R(\Omega_{+} + \Sigma\wedge\omega)~,
\\
\dd\Omega_{-} &= 4 (\tau_{1\, 0}\, \rho  + \tau_{1\, \perp}\wedge\Omega_{-}) - R(\rho - \Sigma\wedge\Omega_{-})
- \tau_{2\, 0}\wedge\Omega_{+} - \tau_{2\, \perp}\wedge\omega~,
\\
\dd\omega &= \tau_{0}\, \Omega_{-} - 3\, \tau_{1\, 0}\, \Omega_{+} + 3 \, \tau_{1\, \perp}\wedge\omega + *\tau_{3\, \perp}
+ R(\Omega_{+} + \Sigma\wedge\omega)~,
\\
\dd\rho &= 4\, \tau_{1\, \perp}\wedge\rho + \dd\Sigma\wedge\Omega_{-} + \Sigma\wedge R(\rho - \Sigma\wedge\Omega_{-})
- \tau_{2\, \perp}\wedge\Omega_{+}~,
\end{align*}
where we have used Lemma \ref{lemma:lrcorner} and decomposed the $G_{2}$ torsion classes with respect to the ACS
\be
\tau_{1}= \sigma\, \tau_{1\, 0} + \tau_{1\, \perp}~, \qquad\qquad \tau_{2}= \sigma\wedge \tau_{2\, 0} + \tau_{2\, \perp}
\qquad\qquad \tau_{3}= \sigma\wedge \tau_{3\, 0} + \tau_{3\, \perp}~.\label{eq:decomptau}
\ee
The condition that  $\tau_3\in \Omega^{3}_{\bf 27}(Y)$, implies that $\tau_{3\, \perp}$ is type $(2,1)+(1,2)$, $\tau_{3\, 0}$
is primitive  and 
\be
\omega\lrcorner\tau_{3\, \perp} = \tau_{3\, 0}\lrcorner\,\Omega_{+}~.
\ee
Similarly, the condition that  $\tau_{2}\in \Omega^{2}_{\bf 14}(Y)$ implies that $\tau_{2\, \perp}$ is primitive  and 
\be
\tau_{2\, \perp}\lrcorner\,\Omega_{-} = \tau_{2\, 0}~.
\ee

We can now compare these relations with the  $SU(3)$ structure equations \eqref{eq:su3om} and \eqref{eq:su3Om}, and deduce formulas for the torsion classes of the ACMS-induced $SU(3)$ structure on $Y$, $W_{i}$, in terms of the $G_{2}$ torsion classes and the flow of the $SU(3)$ structure along $R$.  After a somewhat lengthy computation\footnote{These computations are very similar to those in reference \cite{delaOssa:2014lma}.}, we find 
\begin{equation}\label{eq:Wtau}
\begin{split}
{\rm Re}\,W_{0}&= \frac{2}{3}\, \tau_{0} + \frac{1}{6}\,  \Omega_{-}\lrcorner R\big(\Omega_{+}+ \Sigma\wedge \omega\big)~,
\\[5pt]
{\rm Im}\,W_{0}&= 2\, \tau_{1\,0} - \frac{1}{6}\, \Omega_{+}\lrcorner R\big(\Omega_{+} + \Sigma\wedge \omega\big)~,
\\[5pt]
2\, W_{1} 
& = 4\, \tau_{1\, \perp}    - J(\tau_{2\, 0}) + \dd_{\perp} \Sigma\lrcorner\, \Omega_{-}
+ \omega\lrcorner (\Sigma\wedge R(\omega))
 ~,
\\[5pt]
2\, {\rm Re}\,\theta 
&= 4\, \tau_{1\, \perp}  + \frac{1}{2}\, J(\tau_{2\, 0})  + \frac{1}{2}\, \omega\lrcorner\, R(\Omega_{+})
+\Omega_{-}\lrcorner R(\Sigma\wedge\Omega_{-})
\\[5pt]
{\rm Re}\,W_{2} &=  - \tau_{3\, 0}^{(1,1)}  - {\cal P}\Big( 
\dd_{\perp} \Sigma 
-\omega\lrcorner \big( \Sigma\wedge R(\Omega_{+}) \big)
\Big)^{(1,1)}~,
\\[5pt]
{\rm Im}\,W_{2} &=  - \tau_{2\, \perp}^{(1,1)}- {\cal P}\Big( R(\omega) 
-\omega\lrcorner \big( \Sigma\wedge R(\Omega_{-}) \big)
\Big)^{(1,1)}~,
\\[5pt]
W_{3}&= {\cal P}\Big( J(\tau_{3\, \perp}) + R\big(\Omega_{+} +\Sigma\wedge \omega\big)^{(2,1)+(1,2)}
\Big)~. 
\end{split}
\end{equation}
In these equations, $\cal P$ denotes that the primitive part of the form is taken. 
Two additional identities appear in this computation:
\begin{align}
2\, \omega\lrcorner \dd_{\perp}\Sigma 
&= - \tau_{0} - \Omega_{-}\lrcorner\, R(\Omega_{+})
~,
\\[5pt]
\big(\dd_{\perp}\Sigma\big)^{(2,0)+(0,2)} &= \tau_{3\, 0}{}^{(2,0)+(0,2)}
+ \left(\tau_{1\, \perp} + \frac{1}{2}\, J(\tau_{2\, 0})
+ R(\Sigma) + \frac{1}{2}\, \omega\lrcorner R(\Omega_{+}) \right)\lrcorner\Omega_{-}~.
\end{align}

 Let $\nabla$ be a connection compatible with the $G_2$ structure with intrinsic torsion classes $\{\tau_{i},i = 0, 1, 2, 3\}$.  Equations \eqref{eq:Wtau} imply  that one can construct on $Y$ a connection $\hat\nabla$  with $SU(3)$ holonomy  with intrinsic torsion classes $\{ {\rm Re}(\theta), W_{0}, W_{1},W_{2},W_{3}\}$.
We remark however that this does not necessarily imply that $\nabla\omega = 0$ and $\nabla\Omega = 0$. In fact, one can prove, recalling  $\omega = i_R(\varphi)$,  that $\nabla\omega  = 0$ {\it if and only if} $\nabla\sigma = 0$. 
In section, \ref{sec:hetACMS} we discusss an application of this condition in the context of supersymmetry enhancement in heterotic string theories.

\section{Heterotic $G_{2}$ systems under the ACMS} \label{sec:hetACMS}

In this section we explore the effect of ACMS on a class of minimally supersymmetric compactifications of the heterotic string \cite{Gunaydin:1995ku,Gauntlett:2001ur, 2001math......2142F,Gauntlett:2002sc, Firedrich:2003, Gauntlett:2003cy,Ivanov:2003nd,Lukas:2010mf,Gray:2012md} that have received attention lately due to, on the one hand, their connection to $G_2$ instanton bundles, and on the other hand, their interesting deformation theory \cite{delaOssa:2016ivz,delaOssa:2017pqy,Fiset:2017auc,delaOssa:2018azc,Clarke:2016qtg,Clarke:2020erl}. In particular, we decompose the description of this class of solutions with respect to an ACMS. We will see that this extra structure allows to make contact with string compactifications with enhanced supersymmetry. We use this to write down the necessary and sufficient constraints on the ACMS for supersymmetry enhancement. 

\subsection{Heterotic $G_2$ systems} \label{sec:het}
Let $Y$ be a seven dimensional manifold with a $G_2$ structure $\varphi$ and let $V$ be a vector bundle on $Y$ with connection $A$.   We are interested in the decomposition of ten dimensional heterotic superstring backgrounds on $(Y, V)$ that preserve minimal supersymmetry under the ACMS, i.e. {\it heterotic $G_2$ systems}.  A heterotic $G_2$ system is defined to be the quadruple 
\be
[ (Y, \varphi), (V, A), (TY, \Theta), H]~,
\ee
where
\begin{itemize}
\item $\varphi$ is an {\it integrable} $G_2$ structure on the seven dimensional manifold $Y$, that is $\tau_{2}=0$. In this case 
the structure equations can be written as 
\begin{align}
\dd_7\varphi &= \tau_0 \, \psi + 3\, \tau_1\wedge\varphi + *\tau_3
= i_T(\varphi)~,\label{eq:g2phiInt}
\\
\dd_7\psi&= 4\, \tau_1\wedge\psi = i_T(\psi)
~,\label{eq:g2psiInt}
\end{align}
where $T$ is the totally antisymmetric torsion given by
\be
T(\varphi)= \frac{1}{6}\,\tau_0\, \varphi - \tau_1\lrcorner\, \psi - \tau_3~.\label{eq:torsion}
\ee 
Note that a $G_2$ structure admits a totally antisymmetric torsion if and only if 
 $\dd\psi\in\Omega^5_{{\bf 7}}$, and $T(\varphi)$ is in fact the torsion of the unique metric connection with a totally antisymmetric torsion.

\item $V$ is a gauge bundle with connection $A$ that is an instanton, i.e. its curvature $\cal F$ satisfies
\be
{\cal F}\wedge\psi = 0~.\label{eq:Vinst}
\ee
\item $\Theta$ is a connection on the tangent bundle $TY$ of $Y$ which is also an instanton
\be
{\cal R}(\Theta)\wedge\psi = 0~,\label{eq:TYinst}
\ee
where ${\cal R}(\Theta)$ is the curvature of $\Theta$.
\item $H$ is a three form defined by 
\be
H = \dd_7 B + \frac{\alpha'}{4}\, ({\cal CS}(A) - {\cal CS}(\Theta))~,\label{eq:flux}
\ee
where ${\cal CS}(A)$ is the Chern-Simons form of the connection $A$
\be
{\cal CS}(A) = \tr \left( A\wedge \dd_7 A + \frac{2}{3}\, A^3\right)~,\label{eq:csform}
\ee
with a similar definition for ${\cal CS}(\Theta)$, and  $B$ is the $B$-field.
The fields $H$, $A$, $B$ and $\Theta$ are constrained such that
\be
H = T(\varphi)~,\label{eq:anomaly}
\ee
where $T(\varphi)$ is given in equation \eqref{eq:torsion} and it is the totally antisymmetric torsion of a (unique) connection $\nabla$ compatible with the $G_2$ structure.
\end{itemize}

\subsection{Decomposition of the $G_2$ structure equations with respect to the ACMS}

In section \ref{ssec:decomstructeq} we presented the decomposition of the structure equations of a general $G_{2}$ structure under the ACMS. 
  We do not find it necessary  to write these relations here as for an integrable $G_{2}$, all we need to do is to set $\tau_{2}= 0$.

\subsection{Instanton conditions}\label{ssec:InstConds}

We want to discuss how the instanton conditions are decomposed under the ACMS on $Y$.

Let 
\be
A = \sigma\, a_0 + a~,\label{eq:Adecomp}
\ee
be the composition under the ACMS of the gauge connection $A$.
Then, the curvature $\cal F$ of the connection $A$ decomposes accordingly as
\be
{\cal F} = \sigma\wedge {\cal F}_0 + {\cal F}_\perp~,\label{eq:Fdecomp}
\ee
where
\begin{align}
{\cal F}_0 &= - \dd_a a_0 + R(a + \Sigma\, a_0)~,\label{eq:F0}
\\
{\cal F}_\perp &= {\cal F}(a) + \dd\Sigma\,\, a_0 - \Sigma\wedge R(a+ \Sigma\, a_0)~.
\label{eq:Fperp}
\end{align}
It is not too difficult to show that the instanton condition ${\cal F}\wedge\psi= 0$
for the curvature $\cal F$ is equivalent to the constraints
\begin{align}
\omega\lrcorner {\cal F}_{\perp} &= 0~,\label{eq:primFperp}
\\
{\cal F}_{\perp}\lrcorner \Omega_{-} & = {\cal F}_{0}~.\label{eq:NoholFperp}
\end{align}

It is instructive to note that in terms of
\be
\hat a = a + \Sigma\, a_{0}~, \label{eq:hata}
\ee
the components of the curvature are
\be
{\cal F}_{0}= - \dd_{\hat a}\, a_{0}  + R(\hat a)~,\label{eq:Fcomps}
\qquad\qquad
{\cal F}_{\perp} = {\cal F}(\hat a) - \Sigma\wedge {\cal F}_{0}~,
\ee
and the instanton conditions become
\be
\omega\lrcorner {\cal F}(\hat a) = - {\cal F}_{0}\lrcorner J(\Sigma)~,
\qquad\qquad
{\cal F}(\hat a)\lrcorner\,\Omega_{-} = {\cal F}_{0} + {\cal F}_{0}\lrcorner(\Sigma\lrcorner\,\Omega_{-})~.
\ee
These conditions do not correspond to $\hat a$ instantons on the transverse geometry unless, for example, ${\cal F}_{0}= 0$.  

The component ${\cal F}_0$ of the field strength $\cal F$ can be interpreted as a trivial deformation of the gauge connection $A$ under the one parameter group of diffeomorphisms generated by the vector $R$.  To see this, recall \cite{delaOssa:2017pqy}\footnote{In \cite{delaOssa:2017pqy} it was proven that trivial deformations of the heterotic $G_2$ system are exact in the cohomology of a nilpotent operator.  We refer the reader to this reference for details.} that a trivial deformation of $A$ due to a gauge transformation with parameter $\epsilon$ together with a diffeomorphism generated by a vector $V$ is given by
\[
(\delta A)_{triv}= \dd_A\epsilon + i_V({\cal F})~.
\]
The second term gives the component ${\cal F}_0$ when $V=R$.  In other words,  ${\cal F}_0$ gives the variation of $\hat a$
along the integral curves of $R$ up to a gauge transformation of $\hat a$ generated by $a_0$.  Note that if $R$ is a Killing vector then ${\cal F}_0=0$ and, in this case, $\hat a$ is an $SU(3)$ instanton. We will return to this case in section \ref{sec:N2}.

Similar considerations apply to the instanton connection $\Theta$ on the tangent bundle of $Y$.  In this case we decompose the connection $\Theta$ on the tangent bundle of $Y$ under the ACMS as
\[
\Theta = \sigma\wedge\theta_0 + \theta_\perp~.
\]
The curvature of this connection is thus decomposed as
\[
{\cal R}(\Theta) = \sigma\wedge {\cal R}(\Theta)_0 + {\cal R}(\Theta)_\perp~,
\]
where, in terms of $\hat\theta = \theta + \Sigma\,\theta_0\, $, 
the instanton conditions become
\be
\omega\lrcorner {\cal R}(\hat\theta) = - {\cal R}_{0}\lrcorner J(\Sigma)~,
\qquad\qquad
{\cal R}(\hat\theta)\lrcorner\,\Omega_{-} = {\cal R}_{0} + {\cal R}_{0}\lrcorner(\Sigma\lrcorner\,\Omega_{-})~.
\ee
Just as before, the component ${\cal R}_0$ of the curvature ${\cal R}(\Theta)$ is interpreted as a trivial deformation of the instanton connection $\Theta$  on $TY$ under the one parameter group of diffeomorphisms generated by the vector $R$. When $R$ is a Killing vector, this component vanishes and  $\hat\theta$ is an instanton.

\subsection{Anomaly cancellation condition}
Let
\be
H = \sigma \wedge H_0 + H_\perp~,\label{eq:Hdecomp}
\ee
be the decomposition of the flux $H$ with respect to the ACS.  Then, recalling equations \eqref{eq:flux} and  \eqref{eq:csform}, the terms  in the flux decomposition are given by
\begin{align}
H_0 
 &= - \dd b_0 + R(\hat b)
 + \frac{\alpha'}{4}\, 
 \left(\tr \big( a_0\, \dd\hat a -  \hat a\wedge {\cal F}_0 \big)
 - \tr \big( \theta_0\, \dd\hat\theta -  \hat \theta\wedge {\cal R}_0 \big)
 \right)~, \label{eq:H0}
\\[10pt]
H_\perp 
&= \dd\hat b + \frac{\alpha'}{4}\,
\Big({\cal CS}(\hat a) - {\cal CS}(\hat\theta)\Big)
- \Sigma\wedge H_{0}
~,\label{eq:Hperp}
\end{align}
where
\[
\hat b = B_\perp + \Sigma\wedge B_0~, \qquad b_0 = B_0~.
\]

We can interpret $H_0$ as a trivial deformation of the $B$ field due to the diffeomorphism generated by $R$.  In fact, up to an ambiguity of an exact two form, a trivial deformation of the $B$-field due to a diffeomorphism generated by the a vector $V$ and a gauge transformation with parameter $\epsilon$, is given by \cite{delaOssa:2017pqy}
\be
(\delta B)_{triv} =   i_{V}\, (H) + \frac{\alpha'}{2}\, {\tr}(\epsilon { \cal F })~.
\ee
When $V=R$, the first term is indeed $H_0$ as claimed. If $R$ is an isometry, then $H_0$ vanishes up to an exact two form.   We will see the implications of this result in section \ref{sec:N2} as well as in an example in section \ref{sec:nilman}.

The anomaly cancellation condition is the requirement that the flux $H$  equals the torsion $T(\varphi)$ of the connection with $G_{2}$ holonomy determined uniquely by $\varphi$ (see equations \eqref{eq:torsion} and \eqref{eq:anomaly}).  Under the ACMS we have then
\be
H_{0}= T_{0}({\varphi})~, \qquad H_{\perp} = T_{\perp}(\varphi)~.\label{eq:Tdecomp1}
\ee
where
\be
T = \sigma\wedge T_{0} + T_{\perp}~, \label{eq:Tdecomp}
\ee
and
\begin{align}
T_{0}&= \frac{1}{6}\, \tau_{0}\,\omega - \tau_{1\, \perp}\lrcorner\ \Omega_{-} - \tau_{3\, 0}~,
\label{eq:T0}\\[5pt]
T_{\perp} &= \frac{1}{6}\, \tau_{0}\,\Omega_{+} + \tau_{1\, 0}\, \Omega_{-} +J(\tau_{1\, \perp})\wedge\omega - \tau_{3\, \perp}~.
\label{eq:Tperp}
\end{align}
One can write expressions for $T_{0}$ and $T_{\perp}$ in terms of the torsion classes of $SU(3)$ structure induced by the ACMS using equations  \eqref{eq:Wtau} with $\tau_{2}= 0$.

\subsection{$N=1$ superpotential in terms of the ACMS}

A compactification of the heterotic string on a heterotic $G_2$ system leads to minimally  supersymmetric ($N=1$) effective field theory on either $AdS_3$ or three dimensional Minkowski space time.
It has been shown \cite{de_Ia_Ossa_2020} that, up to an overall constant, the superpotential $W$ of the $N=1$ effective theory is given by
\be
W = \int_Y e^{-2\phi}\ \left((H + h\, \varphi)\wedge\psi 
- \frac{1}{2}\, \dd_7\varphi\wedge\varphi\right)~,\label{eq:superpot}
\ee
where $h$ is the three dimensional flux which is related to the curvature of the three dimensional space-time and $\phi$ is the dilaton field. It was also shown in \cite{de_Ia_Ossa_2020} that this superpotential is a functional of the fields whose critical points give the conditions for preservation of $N=1$ supersymmetry in three dimensions, or equivalently, the conditions defining the $G_2$ structure. 
Furthermore, it was shown  that supersymmetry requires, apart from the constraints described above, that $h$ be proportional to the torsion class $\tau_0$, and that $\tau_1$ is $\dd$-exact, specifically
\[ h = \frac{1}{3}\, \tau_0~,\qquad \tau_1 = \frac{1}{2}\, \dd\phi~.\]

As we have seen in section \ref{sec:ACMS}, the ACMS $(J, R, \sigma, g_\varphi)$ on a manifold $Y$ with a $G_2$ structure $\varphi$, implies that the $G_2$ structure can be decomposed in terms of an underlying transverse $SU(3)$ structure $(\omega_\varphi, \Omega_+)$. This is summarised in Proposition~\ref{prop:restrict}. In a similar way, we may use equations \eqref{eq:g2vol}, \eqref{eq:su3comp},
\eqref{eq:d7varphidecomp}, and \eqref{eq:Hdecomp}, to decompose the various terms in the superpotential \eqref{eq:superpot}. 
A short computation gives
\be \label{eq:wdecomp}
\begin{split}
W 
& = \int_Y e^{-2\phi}\, \sigma\wedge {\rm Im}
\left( \left[
\, \hat H + i\, \dd_{\perp}\omega 
\right.\right.
\\[5pt]
&\qquad\left.\left.
+ \frac{1}{8}\, \left( 
\omega\lrcorner(H_0 - \dd_\perp\Sigma) + 7\, h 
- \left( \Sigma\wedge H_{0} + \frac{i}{2}\, R(\Omega_+)\right)\lrcorner \bar\Omega\,
\right)\, \bar\Omega
\right]\wedge\Omega\right)
~,
\end{split}
\ee
where
\be
\hat H = \dd\hat b 
+ \frac{\alpha'}{4}\, \big({\cal CS}(\hat a) 
- {\cal CS}(\hat\theta )\big)~.
\label{eq:Hhat}
\ee

This decomposition provides links between the heterotic $G_2$ system and the six dimensional Strominger--Hull system \cite{Strominger:1986uh,Hull:1986kz}.\footnote{Such links were discussed from a different perspective in \cite{delaOssa:2017gjq}.} In particular, this is evident from the first two terms in the square bracket in \eqref{eq:wdecomp}, where we recognize the $SU(3)$ superpotential of the latter system.\footnote{The remaining terms of $W$ are related to the non-transverse objects.}   This is consistent with the fact that the systems need not be related by a circle reduction, and that the heterotic $G_2$ system need only preserve half of the supersymmetry of the Strominger--Hull system. Indeed, in contrast to Strominger--Hull system, we cannot identify a holomorphic superpotential. As a further comment we note that, as is true for any four dimensional $N=1$ theory, the Strominger--Hull superpotential captures the F-term constraints, but not the D-terms. In contrast, for three dimensional $N=1$ theories, there is no decomposition into F- and D-terms, and indeed the $G_2$ superpotential captures also the D-term constraints of the Strominger--Hull system.

\subsection{Three dimensional theory with $N=2$} \label{sec:N2}

Consider now the following question:  What are the constraints on the heterotic $G_2$ system to have supersymmetry enhanced to $N=2$? This question was already addressed in \cite{Gran:2005wf,Gran:2007kh,Gran:2016zxk}.  In this section, we approach the question using an almost contact metric structure and show that supersymmetry enhancement is equivalent to the existence of a particular kind of ACMS on $Y$.  

We know that any manifold $Y$ with a $G_2$ structure admits a covariantly constant nowhere vanishing spinor $\eta$.  Given the existence of a nowhere vanishing vector $R$, the manifold $Y$ admits another nowhere vanishing spinor  $R\eta$ which is not, however, necessarily covariantly constant. In this section we deduce the geometric conditions under which the spinor $R\eta$ is covariantly constant too. As we will see in this section, this has interesting applications, as for example, to the construction of three dimensional theories with $N=2$ supersymmetry by constraining further the heterotic $G_2$ system such  that $R\eta$ is indeed covariantly constant.  As we will see, with two covariantly constant spinors at hand, the structure group of $Y$ reduces further to a certain type of $SU(3)$ structure.

The requirement that $\nabla(R\eta) = 0$ is equivalent to the condition that $R$ is itself covariantly constant. 
Equivalently (as the connection $\nabla$ is metric), we require that $\sigma$ is covariantly constant
\be
\nabla_a\sigma_b = 0~.\label{eq:covconstsigma}
\ee
As we remarked earlier, we now have $\nabla\omega = 0$ and 
$\nabla\Omega= 0$ so the holonomy of the $G_2$ compatible connection $\nabla$ is reduced to $SU(3)$.
Symmetrising this equation with respect to the indices $a,b$, we obtain that $R$ {\it must be a Killing vector}.
When $R$ is a Killing vector, $\varphi$ becomes independent of the coordinate $r$ and hence,  the transverse forms $\omega$, $\Omega$, and $\Sigma$
are also independent of $r$.
The antisymmetric part of equation \eqref{eq:covconstsigma} gives a further constraint
\be
\dd_7\sigma = i_T(\sigma) = T_{0} ~.\label{eq:N2cond}
\ee
The authors of \cite{Gran:2005wf, Gran:2007kh} 
obtain precisely condition \eqref{eq:covconstsigma} as part of the requirements to obtain supersymmetry preserving backgrounds.  In fact, a ten dimensional backgound on $AdS_3\times Y$, where $Y$ is a manifold with a $G_2$ structure, demands the existence of an extra covarianly constant one form to have an extra supersymmetry and hence reducing the the holonomy of the $G_2$ connection to $SU(3)$.  

The relations between the $G_2$ torsion classes and the induced $SU(3)$  torsion classes
given in  equations \eqref{eq:WtauSym} greatly simplify because we now have to set $\tau_2= 0$ and, $R$ being a Killing vector, means that
$R(\Sigma) = 0$, $R(\omega) = 0$ and $R(\Omega) = 0$.  We then find
\be\label{eq:WtauSym}\begin{split}
{\rm Re}\,W_{0}&= \frac{2}{3}\, \tau_{0}= - \frac{4}{3}\, \omega\lrcorner\, \dd\Sigma
~,\qquad\qquad {\rm Im}\,W_{0}= 2\, \tau_{1\,0}~,
\\[5pt]
2\, W_{1} & = 4\, \tau_{1\, \perp} + \dd \Sigma\lrcorner\, \Omega_{-} ~,
\qquad\qquad
2\, {\rm Re}\,\theta = 4\, \tau_{1\, \perp} ~,\\[5pt]
{\rm Re}\,W_{2} &=  - \tau_{3\, 0}^{(1,1)}  - {\cal P}\big( \dd \Sigma \big)^{(1,1)}~,
\qquad \qquad {\rm Im}\,W_{2} =  - \tau_{2\, \perp}^{(1,1)}~,
\\[5pt]
W_{3}&= {\cal P}\big( J(\tau_{3\, \perp}) \big)~,
\\[5pt]
\big(\dd\Sigma\big)^{(2,0)+(0,2)} &= \tau_{3\, 0}{}^{(2,0)+(0,2)}
+ \tau_{1\, \perp}\lrcorner\Omega_{-}~. 
\end{split}\ee

As discussed above, a Killing vector is necessary but not sufficient for an enhancement of the supersymmetry, we need furthermore to impose the condition \eqref{eq:N2cond}. 
Using \eqref{eq:WtauSym}  in  \eqref{eq:T0}, the condition \eqref{eq:N2cond} becomes
\be
\dd\Sigma = T_0 = - \frac{1}{3}\, (\omega\lrcorner\dd\Sigma)\, \omega + {\rm Re} W_2 + {\cal P}(\dd\Sigma)^{(1,1)} - (\dd\Sigma)^{(2,0)+(0,2)}~.\label{eq:N2condBis}
\ee
Therefore $\dd\Sigma$ must be a primitive $(1,1)$ form and ${\rm Re} W_2 = 0$. Moreover, the fact that
$\omega\lrcorner\dd\Sigma= 0$ means that ${\rm Re} W_0 = 0$, that is $\tau_0 = 0$. 
For the heterotic string we need to add the constraint that $2\, \tau_1 = \dd_7\phi$.  This implies that %
\be
\tau_{1\, 0} = 0~, \qquad 2\,\tau_{1\, \perp} = \dd \phi~.
\ee
Therefore
\be
{\rm Im}\, W_0= 0~.
\ee

In summary, the torsion classes of the $SU(3)$ on the transverse geometry satisfy
\begin{align*}
W_{0}&=  0~,\qquad  W_{2} = 0~,
\\[5pt]
W_{1} &={\rm Re}\,\theta = \dd\phi = -  \tau_{3\, 0}\lrcorner\, \Omega_{-}~,\qquad
\\[5pt]
W_{3}&= {\cal P}\big( J(\tau_{3\, \perp}) \big)
~, 
\end{align*}
and we also have
\[
\tau_0 = 0~, \qquad \dd\Sigma = - \tau_{3\, 0}^{(1,1)}~,
\]
 where $\tau_{3\,0}$ is primitive.  
 
 As an example, we note that, if the manifold $Y$ has $G_2$ holonomy, then the transverse bundle ${\rm Ker}(\sigma)$ is necessarily integrable, so  $Y$ is a codimenion one foliation with leaves which are Calabi--Yau three folds.

More generally, the vanishing of $W_0$ and $W_2$  means that the almost complex structure  $J$ on the transverse geometry is integrable,
and the fact that $W_1$ (and hence ${\rm Re}\,\theta$) are exact imply that the transverse geometry is conformally balanced.
Therefore the transverse geometry has the $SU(3)$ structure of the Strominger-Hull system. 
Moreover, the fact that $\dd\Sigma$ is a primitive $(1,1)$ form means that the manifold $Y$ has the structure of a $U(1)$ principal bundle with a holomorphic connection $\Sigma$ over the transverse geometry.
Finally, we note that the vanishing of $\tau_0$ implies that the three dimensional spacetime is Minkowski space.

Consider now the instanton conditions.  Following the discussion at the end of section \ref{ssec:InstConds}, 
we recall that ${\cal F}_0$ represents the change of the connection $A$ with respect to the vector  $R$.  As $R$ is, in our case, a Killing vector, we have ${\cal F}_0 = 0$ and it follows that $\hat a$ is a holomorphic instanton on the transverse geometry.  Similarly, ${\cal R}_0 = 0$ and $\hat\theta$ is also an instanton.

For the {anomaly} cancellation condition, we earlier saw that if $R$ is a Killing vector, then $H_0$ must be an exact two form, a conclusion that was arrived at by studying the symmetries of the heterotic system.  Interestingly, the condition that $H_0$ be an exact form is precisely the content of equation \eqref{eq:N2condBis}, which fixes this exact form to be  
\[
H_0 = T_0 = \dd\Sigma~,
\]
where we recall that the second equality comes from the fact that $R$ is not just a Killing vector, but is also covariantly constant. 
For completeness we note that the transverse part of the flux becomes 
\[
H_\perp = - \Sigma\wedge\dd\Sigma + \hat H~,
\]
where $\hat H$ is defined in  equation \eqref{eq:Hhat}.  Note, furthermore, that the Bianchi identity of the anomaly cancellation condition becomes
\be
\dd H_\perp = - \dd\Sigma\wedge\dd\Sigma + \frac{\alpha'}{4} \Big( \tr ({\cal F}(\hat a)\wedge {\cal F}(\hat a)) 
- \tr ({\cal R}(\hat\theta)\wedge {\cal R}(\hat \theta)) \Big)~.
\ee

We remark that it has not been necessary to require the integrability of  ${\rm Ker}(\sigma)$.
That is,  it is not necessary to require that $\dd\Sigma = 0$ to have $N=2$ in three dimensions, as one might have expected.

\section {$SU(2)$ structures on manifolds with a $G_2$ structure}\label{sec:ACM3S}

The  $SU(3)$ structure discussed in the previous section is not the only ``bonus'' restriction on the topology of $G_2$ structure manifolds. Indeed, there are {\emph{two}} non-vanishing vector fields on a manifold $Y$ with a $G_2$ structure \cite{10.2307/1970439,thomas1969}, which may be combined with the  $G_2$ compatible spinor to form two additional nowhere-vanishing spinors. These three spinors reduce the structure group of the manifold to $SU(2)$ \cite{friedrich1997nearly}. Moreover, the existence of two well-defined vectors implies that there is an almost contact metric 3-structure (ACM3S) on $Y$ (that is a reduction of the structure group to $ {\bf 1}_3  \times Sp(1)$) \cite{kuo1970} (see also \cite{Todd:2015era}). In this section, we will expand on this topic, and show how the $Sp(1) \cong SU(2)$ structure is embedded in the $G_2$ structure. As in the preceding section, we will describe the ACM3S in terms of differential forms. We will also clarify when the ACM3S leads to a involutive decomposition of the tangent bundle $TY$, and expand on the relation to associative and coassociative submanifolds. Finally, we will study the space of ACM3S.

\subsection{Almost contact 3-structure}

It is a classical result by E.~Thomas, that any compact, orientable 7-dimensional manifold $Y$  admits two globally defined, everywhere linearly independent vector fields $R^1,R^2 \in \Gamma(TY)$ \cite{10.2307/1970439,thomas1969}.  Suppose now that $Y$ has $G_2$ structure $\varphi$ with metric $g_\varphi$.  We may then assume, without loss of generality, that the 2-frame $(R^1,R^2)$ consists of vectors that are orthonormal. Indeed, given two non-orthogonal, but linearly independent, vectors $(R^1,R^2)$, we can form two orthogonal ones using the $G_2$ cross product: $(R^1, R^1 \times_{\varphi} R^2)$. We can also normalise the vectors by dividing by their norms as calculated with the $G_2$ metric. Thus any $G_2$ manifolds allow orthonormal 2-frames of vectors. 

Each vector $R^\alpha$ will be associated to an ACMS, in the sense discussed in section \ref{ssec:g2ACS}.  We thus have \cite{Todd:2015era} (see also \cite{Arikan:2012acs,Arikan:2011acs}) two dual one-forms $\sigma^{\alpha}$ so that $(J^{\alpha}, R^{\alpha}, \sigma^{\alpha}, g_{\varphi})$, for $\alpha = 1,2$, are two ACMS on $Y$.  If the structures furthermore satisfy the relations
\be
\label{eq:ac3s}
\begin{split}
&\sigma^1 (R^2) = \sigma^2 (R^1) = 0 \\
&J^1 (R^2) = - J^2 (R^1) \\
&\sigma^1 \circ J^2 = - \sigma^2 \circ J^1 \\
&J^1 J^2 - R^1 \otimes \sigma^2  = -J^2 J^1 +   R^2 \otimes \sigma^1\; ,
\end{split}
\ee
Kuo \cite{kuo1970} has shown that $Y$ admits a third ACMS given by
\be
J^3 = J^1 J^2 - R^1 \otimes \sigma^2  \; , \; R^3 = J^1 (R^2) \; , \; \sigma^3 = \sigma^1 \circ J^2 \; .
\ee
Together, these three ACMS then satisfy \eqref{eq:ac3sintro}, and hence define an almost contact 3-structure (AC3S) \cite{kuo1970}. 
In fact, one can show that, for two ACS associated to the same metric, the last constraint in \eqref{eq:ac3s} implies the other three constraints \cite{kuo1970}. 

From the definition of an AC3S, we have the following useful identities
\be
\sigma^{\alpha}(R^{\beta}) = \delta^{\alpha \beta} \; \; , \; \; J^{\alpha} (R^{\beta}) = \epsilon^{\alpha \beta \gamma} R^{\gamma}
\; \; , \; \;  \sigma^{\alpha} \circ J^{\beta} = - \sigma^{\beta} \circ J^{\alpha}
\; .
\ee
In addition, %
\be
R^{3\, a} =[J^1(R^2)]^a =  \varphi^a{}_{bc}R^{1\, b} R^{2\, c}
\; \mbox{ i.e. } \; 
R^3 = R^1 \times_{\varphi} R^2 \label{eq:SU2-R3}
\ee
or, equivalently,
\be \label{eq:useful}
\sigma^{\gamma} = \frac{1}{2}\epsilon^{\alpha \beta \gamma} i_{R^{\beta}} i_{R^{\alpha}} \varphi \; ,
\ee
which will be used below when we discuss the $SU(2)$ decomposition of the $G_2$ structure.

Finally, it was recently proven by Todd \cite{Todd:2015era}  that on any $G_2$ structure manifold,  two ACMS's $(J^{\alpha}, R^{\alpha}, \sigma^{\alpha}, g_{\varphi})$ will automatically satisfy the last constraint of  \eqref{eq:ac3s}. Thus, we have
\begin{theorem} [Todd, \cite{Todd:2015era}] Let $(Y,\varphi)$ be a compact and boundary-less 7-manifold with  $G_2$ structure $\varphi$. Then $Y$ admits an almost contact metric 3-structure which is compatible with the G2-metric. 
\end{theorem}

Note that the three ACMS have different contact forms $\sigma^{\alpha}$, but the fact that they are associated to the same $G_2$ metric,
\be
g_{\varphi}(J^{\alpha} u, J^{\alpha} v) = g(u,v) - \sigma^{\alpha}(u) \sigma^{\alpha}(v) \; \; , \; \;  {\alpha}=1,2,3 \; \; {\rm (no \; sum)}\; , 
\ee
implies the resulting almost contact 3-structure is indeed an almost contact metric 3-structure (ACM3S). %

We observe, in particular, that the ACM3S involves fixing a distinguished orthonormal three-frame, $(R^1,R^2,R^3)$, which induces a decomposition of the tangent bundle
\begin{equation}
TY\cong \cT\oplus \cT^\perp,\label{eq:TMdecompSU2}
\end{equation} 
where $\cT$ is trivial rank three bundle with a distinguished trivialisation induced by the three-frame $(R^1,R^2,R^3)$. The second factor, $\cT^\perp$, is then the orthogonal complement.  With this data, we are able to identify $\cT$ as a trivial bundle of imaginary quaternions, with a product induced by the $G_2$ cross product. This follows immediately from \eqref{eq:SU2-R3}. 

The choice of ACM3S therefore reduces our structure group $G_2\rightarrow {\bf 1}_3\times H,$ for some $H\subset G_2$. In fact, results of Kuo, \cite{kuo1970}, show that $H=Sp(1)\cong SU(2)$. {One way to see this, from the $G_2$ perspective, is to observe that the subgroup of $G_2$ that preserves three orthonormal vectors is, indeed, $SU(2)$, from which the result follows.} Note, in particular, that the rank four bundle, $\cT^\perp$ has reduced structure group $SO(4)\rightarrow SU(2)$.  A reduction of structure group leads to distinguished differential forms living in irreducible representations of the reduced group and we will explicitly exhibit the forms induced by the reduction of $\cT^\perp$'s structure group.  These structure forms will be familiar to readers that have studied four dimensional $SU(2)$ structure manifolds, but it is important to bear in mind that $\cT^\perp$  need not be tangent to any underlying four manifold.\footnote{The conditions that $\cT^\perp$ must satisfy in order for such manifolds to exist are reviewed in section \ref{sec:su2int}.} This decomposition is purely at the level of the bundle.

The splitting of the tangent bundle, \eqref{eq:TMdecompSU2} induces an analogous decomposition of the cotangent bundle 
 \be
T^*Y = \cT^* \oplus \cT^{*\, \perp} \equiv {\rm Span}\{\sigma^1,\sigma^2,\sigma^3\} \oplus {\rm Span}\{\sigma^1,\sigma^2,\sigma^3\}^{\perp} \; ,\label{eq:cotandecomp}
\ee
and, consequently, a decomposition of the differential forms. In particular, we will refer to a $k$-form, $\lambda$, as being {\it $\alpha$-transverse} if it satisfies 
 \be
 i_{R^{\alpha}} (\lambda) = 0 \; .
 \ee
 More generally, an arbitrary form $k$-form, $\lambda$, can be decomposed as
\be
\lambda = \sum_{\alpha} \sigma^{\alpha} \wedge\lambda_0^{\alpha} +   \lambda_\perp~,
\ee
where
\[
i_{R^{\alpha}} (\lambda) = \lambda_0^{\alpha} ~,\qquad i_{R^{\alpha}} (\lambda_0^{\alpha} ) = 0~, \quad \mbox{ and, }  \; i_{R^{\alpha}} (\lambda_\perp) = 0\,,\;\forall  \alpha~ .
\]

Recall that each ACS has an associated fundamental two-form, \eqref{eq:iRphi}, 
\be \label{eq:omalpha}
\omega^{\alpha} = i_{R^{\alpha}}(\varphi)\,.
\ee
Evidently, $\omega^{1}$ is a $1$-transverse two-form, but it is neither $2$- nor $3$-transverse; as a consequence it is not in $\Gamma(\Lambda^2\cT^{* \perp})$, but is instead a linear combination
 \be
 \omega^{\alpha} = \sum_{\beta \neq \alpha} \sigma^{\beta} \wedge \omega_{0 \, \beta}^{\alpha} +    \omega^{\alpha}_{\perp} \; .
 \ee
where we recall that $\omega^{\alpha}_{\perp}$ is, in fact, transverse with respect to all three ACMS.  
Moreover, \eqref{eq:useful} implies that e.g.
\be
\sigma^3 = i_{R^2}  \omega^{1} = -i_{R^1}  \omega^{2}
\ee
and we thus have that $\omega_{0 \, 1}^{2} = -\sigma^{3}$, or, in general, 
 \be \label{eq:omegaac3s}
 \omega^{\alpha} =  
 \frac{1}{2}\epsilon^{\alpha \beta \gamma} \sigma^{\beta} \w \sigma^{\gamma}
 +    \omega^{\alpha}_{\perp} \; .
 \ee
 
Next, we decompose the $G_2$ structure form, $\varphi$ with respect to the ACM3S.  Using that the trivial bundle, $\cT\subset TY$, can be interpreted, fibrewise, as the imaginary quaternions sitting inside the imaginary octonions, we can quickly deduce that $\varphi_\perp=0$. This is because the octonionic product of any two elements in $\Ima(\IH)^\perp\subset \Ima(\IO)$ will necessary land back in $\Ima(\IH)$, while $\varphi_\perp$ projects this product back into the complement $\Ima(\IH)^\perp$.
 
 More concretely, we can consider a local frame, $e^i$, such that $e^{4+\alpha}:=R^\alpha$, and $e^1,e^2,e^3,e^4$ are chosen such that $\varphi$ takes the standard form
 \be \label{eq:phistandard}
\varphi_0 = (e^{12}+e^{34}+e^{56})\w e^7 + e^{135}-e^{146}-e^{236}-e^{245}\,.
\ee 
 In such a local frame, $\varphi_\perp$ will be those terms of $\varphi_0$ in which none of $e^5,e^6,e^7$ appear and one observes that there are no such terms.
 
Furthermore, comparing the terms that appear in \eqref{eq:phistandard} with the expansion \eqref{eq:omegaac3s}, we can explicitly expand $\varphi$:
 \be
 \varphi=\frac{1}{3!} \epsilon_{\alpha \beta \gamma}  \sigma^{\alpha} \w \sigma^{\beta} \w \sigma^{\gamma}
+ \sum_{\alpha}\sigma^{\alpha} \wedge \omega^{\alpha}_{\perp} \; ,\label{eq:su2decomp-phi}
 \ee

 The expansion of $\psi$ can be straightforwardly computed, using either a local frame and its standard form, or applying the Hodge star. Either way, one obtains
\be \label{eq:su2decomp-psi}
\psi = \dd {\rm vol}_\perp + \frac{1}{2}\epsilon_{\alpha \beta \gamma} \sigma^{\alpha} \w \sigma^{\beta} \w \omega^{\gamma}_\perp \; .
\ee

Note that we use $\dd {\rm vol}_\perp$ to refer to the canonical section of $\Lambda^4(\cT^{\perp,*})$, although there may not be a four dimensional manifold, even locally, for which $\dd{\rm vol}_\perp$ is a volume form. Whenever $\cT^\perp$ is integrable (see the next subsection), then $\dd {\rm vol}_\perp$ is indeed the volume form for the leaves of the corresponding foliation.

In the preceding, we have come across a triple of real two-forms $\omega^{\alpha}_{\perp}$, $\alpha=1,2,3$. These characterise the reduced $SU(2)$ structure of the rank four bundle, $\cT^\perp$.  This will be familiar to readers comfortable with $SU(2)$ structures in dimension four and we briefly review this setting in Appendix \ref{sec:su2}.  To be sure that these really are the correct differential forms, we must check that
\be
\omega^{\alpha}_{\perp} \w \omega^{\beta}_{\perp}
=
2 \delta^{\alpha \beta} \dd {\rm vol}_\perp \; .
\ee
That this holds follows directly from the decompositions \eqref{eq:su2decomp-psi} \eqref{eq:su2decomp-phi}. Indeed, on the one hand, we have
\be
0 \neq 7 \dd {\rm vol}_\varphi = \varphi \w \psi = \sigma^1 \w \sigma^2 \w \sigma^3 \w (\dd {\rm vol}_\perp + 3  \omega^{\alpha}_{\perp} \w \omega^{\alpha}_{\perp}) \;,
\ee
showing that $\omega^{\alpha}_{\perp} \w \omega^{\alpha}_{\perp}
= 2 \dd {\rm vol}_\perp$. On the other hand, we have
\be
0 = \varphi \w \varphi \sim \sigma^{\alpha} \w \sigma^{\beta} \w \omega^{\alpha}_{\perp} \w \omega^{\beta}_{\perp}
\ee
so $\omega^{\alpha}_{\perp} \w \omega^{\beta}_{\perp}$ must vanish when $\alpha \neq \beta$.

We now return to the role that the unit vector fields, $R^\alpha$ played in the above computations.  In particular, it is important that fixing the splitting \eqref{eq:TMdecompSU2} does not fix the ACM3S. The extra data required is a global identification of $\cT$ with the imaginary quaternions, $\Ima(\IH)$, along with the multiplication induced by $\varphi$. This is equivalent to a choice of three orthonormal vector fields, satisfying \eqref{eq:SU2-R3}.  In particular, after a choice of splitting, \eqref{eq:TMdecompSU2}, there remains an interesting space of compatible almost contact metric 3-structures, which we study in Subsection \ref{sec:count-ac3s}. To the authors' knowledge, this has not been explored in the literature and we provide some initial results below. 

In contrast, a decomposition $TY\cong \cT_1\oplus\cT_1^\perp$ into a trivial one-dimensional bundle, $\cT_1$, and its orthogonal complement, $\cT^\perp_1$, has a much less interesting space of almost contact structures. Indeed, recall from Section \ref{sec:ACMS} that the data needed for an ACMS on a $G_2$ structure manifold is precisely that of a unit vector field.\footnote{Just as an ACM3S allows us to identify $\cT$ with the imaginary quaternions, an ACMS allows us to identify $\cT_1$ with the imaginary complex numbers.}  The rank one bundle, $\cT_1$ admits precisely two such vector fields, say $R$ and $-R$. Choosing either, say $R$ for concreteness, induces an ACMS such that $\cT_1={\rm Span}(R)$ and $\cT_1={\rm Ker}\sigma$. 

To further contrast with the ACMSs of Section \ref{sec:ACMS}, the trivial subbundle, $\cT$, is no longer guaranteed to be tangent to a foliation, i.e. it need not be an integrable distribution.  As a consequence, we can not generally expect to find adapted coordinates in the same vein as \eqref{eq:coords}. In other words, we can not expect there to be even a local product structure of the geometry. We turn to this question now.

\subsection{Integrability}\label{sec:su2int}

In this subsection we will investigate the conditions for the distributions $\cT$ and $\cT^\perp$ to be tangent to a foliation of the seven manifold, $Y$. We will begin with the trivial bundle, $\cT$. It is  a standard result in the study of foliations that this is true if and only if $\cT$  is involutive, for which it suffices that the Lie bracket of the distinguished vector fields, $R^\alpha$, closes. That is, there are real functions $f^{\alpha\beta}_{\;\;\;\gamma}$, $\alpha,\beta,\gamma\in\{1,2,3\}$ such that
\begin{equation}
    [R^\alpha,R^\beta]_x=f^{\alpha\beta}_{\;\;\;\gamma}(x)R^\gamma_x\quad\;\forall\,x\in Y\,.\label{eq:Tinvol}
\end{equation}

Observe that the analagous condition on the rank one bundle induced by an ACMS, ${\rm Span}(R)$, is trivially satisfied since $[R,R]=0$. This is a reflection of the well-known fact that vector fields always admit integrable curves or, to put it another way, that ordinary differential equations always have locally unique solutions. The possible failure of integrability of a higher rank foliation is related to the comparative subtlety of partial differential equations.

Note also that \eqref{eq:Tinvol} is independent of the distinguished vector fields, $R^\alpha$, which can be checked by straightforward computation. In particular, integrability is a property of the subbundle, not of its trivialisation.

Condition \eqref{eq:Tinvol} can be equivalently formulated in terms of the kernel $\cT^{*,\perp}$. Recalling that a one-form, $\xi$ is in $\Gamma(\cT^{*,\perp})$ if and only if $i_{R^\alpha}\xi=0$ for each $\alpha=1,2,3$, then, $\cT$ is involutive if and only if $\dd(\cT^{*,\perp})\subset \cT^{*,\perp}\otimes\Omega^1(Y)$, i.e. that $i_{R^\alpha}i_{R^\beta} \dd\xi=0$ for all $\xi\in\Gamma(\cT^{*,\perp})$.

We can now review the conditions under which $\cT^\perp$ is integrable. Since the presentation of $\cT^\perp$ is not as convenient as that of $\cT$, it is the second of the above characterisations that is most convenient, viz. $\cT^\perp$ is integrable if and only $\dd(\cT^*)\subset \cT^{*}\otimes\Omega^1(Y)$. Using that $\cT^*={\rm Span}\{\sigma^1,\sigma^2,\sigma^3\}$, we conclude that $\cT^\perp$ is involutive if and only if $\dd\sigma^\alpha=\sigma^\beta\w\mu_\beta^\alpha$, for one-forms $\mu_\beta^\alpha$. 
This is the ACM3S analogue of the condition \eqref{eq:dualfrob} for ACMS's.

We can now ask about the relation between integrability of $\cT$, respectively $\cT^\perp$, and the $G_2$ structure on $Y$.  Let us begin by assuming that $\cT$ is integrable, so that $Y$ has a three dimensional foliation with leaves, say $\cL_x$, such that $T_y\cL_x=\cT_x$. Here, $x$ is any point on the leaf, so there is a huge degeneracy in this labelling and, unless it will lead to confusion, we will omit recording this point. We observe that, by construction, $\cL$ has a volume form which is, at each point $y\in\cL_x$, the wedge of the dual one-forms $ \dd{\rm vol}_{\cL,y}=(\sigma^1\w\sigma^2\w\sigma^3)_y$.  Further, this wedge product is nothing but the $G_2$ structure three-form, $\varphi$, restricted to $T_y\cL$, as can be easily seen by choosing a local frame extending $\sigma^\alpha$, cf. \eqref{eq:su2decomp-phi}.  This is precisely the condition that $\cL$ be an associative manifold, \cite{harvey1982calibrated,mclean1998deformations}.

 We have just shown that $\cT$ is integrable if and only if $Y$ is foliated by associative submanifolds. In the case that our $G_2$ structure is closed,  $\dd\varphi=0$, then $\varphi$ is a calibration and the associative submanifolds are the corresponding calibrated cycles, which minimize volume within their homology class. In string theoretic compactification scenarios, calibrated cycles contribute to non-perturbative effects in the effective field theory and it is a long-standing open problem to properly account for these contributions, see \cite{Harvey:1999as,braun2018infinitely,Acharya:2018nbo,Braun:2017uku} for instance, or \cite{Eckhard:2018raj} for a brane world-sheet perspective.  For more general $G_2$ structures, in particular for non-closed $G_2$ structures, the positive three-form $\varphi$ is longer a calibration, the volume-minimizing cycles and corresponding non-perturbative effects are, in general, less understood \cite{Joyce:2016fij}.  The surprising link between three-structures and associatives may offer a tool to be applied to both of these problems and we comment on this possibility in the conclusions, Section \ref{sec:conc}.  
 
At any rate, the condition that $\cT$ be integrable is clearly a strong one. As a final comment on this property, we remark that the leaf, $\cL$, is evidently parallelisable and, once we fix the ACM3S, is equipped with a canonical trivialisation. In three dimensions, every oriented manifold is parallelisable, so this does not add extra conditions on the topology of the leaf.

The case for $\cT^\perp$ is similar. Were $\cT^\perp$ to be integrable, $Y$ would have a four-dimensional foliation, whose leaves will be denoted $\cL^\perp$, $T_x\cL^\perp=\cT^\perp_x$. Choosing a local orthonormal frame extending $(R^1,R^2,R^3)$, say with $e^1,e^2,e^3,e^4$ then we have that in this frame the volume of $\cL^\perp$ will be given by $\dd{\rm vol}_x(\cL^\perp)=e^1\w e^2\w e^3\w e^4$, which is precisely the restriction of the coassociative four form. Submanifolds of a $G_2$ structure manifold with volume form given by the restriction of the $G_2$ structure four form, $\psi$, are called coassociative submanifolds. When $\psi$ is closed, it is a calibration of the seven manifold and coassociative submanifolds are the corresponding calibrated manifolds. Therefore, much of what we said on the interest in associative three-cycles can be said, {\it mutatis mutandis}, for coassociatives. 

 It is interesting to note that {the setting where $\cT^\perp$ is  integrable is reminiscent of the study of $G_2$ manifolds fibred by coassociative cycles which has been promoted by Donaldson \cite{2016arXiv160308391D}, Baraglia \cite{2010JGP....60.1903B} and  Kovalev \cite{2005math.....11150K}, among others. K3-fibrations are also relevant in M-theory/heterotic duality and } have been utilised in defining a conjectural generalisation of mirror symmetry to the seven dimensional case \cite{gukov2003duality,braun2018towards,braun2017mirror}.  It would be interesting to explore how {these topics} interact with the ACM3Ss.

 \subsection{Space of ACM3S}\label{sec:count-ac3s}

 We now return to the study of the space of ACM3 structures compatible with a given $G_2$ structure on a seven manifold, $(Y,\varphi)$. We will denote this space by $\scC$, or $\scC(Y,\varphi)$ if context makes it necessary to emphasise the $G_2$ structure manifold. This is an interesting space in its own right, but it is also  necessary to understand from a physics perspective, since it is far from clear that different choices of ACM3S would yield equivalent effective field theory descriptions of physics. 
 In other words, there may be certain choices of ACM3S that  
provide better descriptions of the low-energy physics.  At the moment, the precise meaning of these choices for physics are unclear and we will simply give a mathematical description of this space.  It seems likely that physics considerations will refine this study and we will comment on these possibilities in the conclusion, see section \ref{sec:conc}. Mathematically, this space, $\scC$, can be seen as a very coarse invariant of the $G_2$ structure, essentially depending only on the topological class of the associated $G_2$ bundle.
 
  One indication of this interdependence is the way parallelisability is reflected by ACM3Ss. This connection comes about as follows: consider any two splittings $\cT\oplus\cT^\perp\cong TY\cong \cT'\oplus\cT^{\prime,\perp}$. Observe that the rank of the sum $\cT + \cT'$ must have at least one fibre (and therefore, also all fibres in an open neighbourhood) with dimension greater than three in order that the splittings be different. On the other hand, if \emph{all} the fibres have rank greater than three, then the manifold is in fact parallelizable. Indeed, since we assume $\cT+\cT'$ is everywhere at least rank 4, then we can conclude that there is at least four orthonormal vector fields, say $R^\mu,\,\mu=1,2,3,4$. Regarding these as unit, imaginary octonions with the $G_2$ cross product, we then have the basic fact that any four orthogonal, imaginary octonions generate the space of all imaginary octonions.   Turning this statement around: whenever the underlying $G_2$ structure manifold is \emph{not} parallelizable, any two splittings will overlap in at least one point.
 
 We will now give a concrete expression for the space of all ACM3Ss induced by the $G_2$ structure. We have seen that these are in one-to-one correspondence with orthonormal three-frames satisfying \eqref{eq:useful}, i.e.~that $R^3=R^1\times_\varphi R^2$. Since this implies that the third orthonormal vector is completely fixed by the first two, we conclude that $\scC$ is simply the space of all orthonormal, ordered pairs of vector fields.  Fibrewise, the space of orthonormal pairs of vectors inside $T_xY\cong \IR^7$ is a so-called Stiefel manifold which we can denote $V_2(T_xY)$. This space has a description as a homogeneous space $G_2/SU(2)$ \cite{harvey1982calibrated}, expressing the fact that $G_2$ acts transitively on orthonomal pairs of vectors, with stabiliser $SU(2)$.  
 Globally, there is a fibre bundle associated to the tangent bundle with typical fibre $V_2(\IR^7)$. We will denote this bundle by $\cV_2(TY)$. A section of $\cV_2(TM)$ is the same as an orthonormal two-frame or, equivalent in our situation, an ACM3S, and therefore the space of ACM3S is the space of sections of $\cV_2(TY)$, $\scC=\Gamma(Y,\cV_2(TY))$.\footnote{Thomas' proof that any $G_2$ structure manifold admits a 2-vector field, \cite{10.2307/1970439, thomas1969} used obstruction theory to show this space of sections is non-empty.}  This space may have non-trivial topology, including non-trivial homotopy groups, which we will investigate in some examples, see Section \ref{sec:examples}.
 
This space of sections, $\scC$,  has a locally trivial fibre bundle structure, where the base corresponds to the space of splittings of the form \eqref{eq:TMdecompSU2} and the fibres are the trivialisations of this bundle.  The projection map is simply taking the span fibrewise. We will now make this more concrete.

The space of decompositions, $TY\cong \cT\oplus\cT^\perp$, with $\cT$ a trivial bundle with associative fibres is, fibrewise, equivalent to choosing a three-plane in the seven-dimensional tangent space with a further constraint that enables it to be regarded as a copy of the imaginary quaternions.  The space of such choices at a given point is the so-called associative Grassmanian, $G(\varphi_x)$, and is, like $V_2(T_xY)$, a homogeneous space with $G_2$ total space,   $G(\varphi_x)=G_2/SO(4)$, \cite{harvey1982calibrated}.  Once again, we can consider a fibre bundle associated to $TY$, now with typical fibre $G(\varphi_0)$. We will call this fibre bundle $\cG(\varphi)\rightarrow Y$ and let $\tilde{\scS}:=\Gamma(Y,\cG(\varphi))$ be the space of sections.  By construction, a section of this bundle corresponds to a rank three subbundle of $TY$ with associative fibres, and thus a splitting $TY\cong\cT\oplus\cT^\perp$, but it is not obvious that the bundle $\cT$ must be trivial. Thus, $\tilde{\scS}$ is not precisely the space we need, it is too big. The subspace of trivial bundles, however, consists of a union of path-connected components of $\tilde{\scS}$. Indeed, the bundle associated to a section in the same path component as that of a trivial bundle is, by definition, homotopic to the trivial bundle and consequentially isomorphic. Therefore, the relevant subspace of all sections will be a disjoint union of certain disconnected components of the space of sections, which we will denote by $\scS\subset\tilde{\scS}$. 
 
We can now look at the fibres of the projection $\scC\rightarrow\scS$, which consists of the the space of orthonormal, associative trivialisations of the corresponding trivial rank three bundle. Since we assume this is not an empty space, we can fix an initial ACM3S by making an arbitrary choice of orthonormal vector fields, $(R^1,R^2,R^3)$ satisfying \eqref{eq:SU2-R3}, i.e. $R^3=R^1\times_\varphi R^2$. 
 
 Any other trivialisation of $\cT$ is given by an orthonormal framing, $(S^1,S^2,S^3)$, and it follows that there is a unique $\Theta\in \Maps(Y,O(3))$ such that each $S^\alpha=\Theta^\alpha_\beta R^\beta$. Imposing that $(S^1,S^2,S^3)$ satisfies condition \eqref{eq:SU2-R3}, implies that $\Theta$ in fact takes values in $SO(3)$. 

We see, then, that the space of orthonormal trivialisations of $\cT$, compatible with \eqref{eq:SU2-R3}, has a free, transitive action of the topological group $\Maps(Y,SO(3))$; in other words it is a torsor for this group. After making an arbitrary choice of basepoint in $Y$, this group factorises into a product, with one factor the space of constant $SO(3)$-valued maps and the other the basepoint-preserving maps. The first factor can be seen as an overall rotation of the vector fields and it seems best to regard this as a redundancy, which we will quotient out. More precisely, we make a choice of basepoint, $x_0\in Y$, which gives us a canonical homomorphism $\Maps(Y,SO(3))\cong\Maps_*(Y,SO(3))\times SO(3)$ where $\Maps_*$ denotes those maps which send the basepoint to the identity, $x_0\mapsto 1\in SO(3)$. Explicitly, this map is given by $\Theta\mapsto ((\Theta\cdot\Theta(x_0)^{-1}),\Theta(x_0))$ and can be interpreted as using a global rotation to ensure that any given ACM3S agrees with that induced by $(R^1,R^2,R^3)$ at $x_0$. 

We will denote this space of trivialisations by $\scT$, which is manifestly independent of the underlying splitting, $s\in\scS$. We will now show that the projection $p:\scC\rightarrow\scS$ is indeed locally trivial.  This argument is reminiscent of the standard construction of the universal bundle over the classifying space $BU(n)$, see for instance \cite{hatcher2016vector, Magill1103965}. 

The topology that we take for both spaces is that induced by compact-open topology on the space of all maps. Focusing on the total space, $\scC$, for concreteness, the compact-open topology has a subbase given by the sets $V(K,U)=\{f:Y\rightarrow \cV_2(TY)\,|\,f(K)\subset U\}$ for $K\subset Y$ compact and $U\subset \cV_2(TY)$ open. An open set in the space of sections will be the intersection of an open set on the space of all maps, with the set of intersections.  The same definitions apply, {\it mutatis mutandis}, for the base space, $\scS$. In particular, for an arbitrary section $s\in\scS$, we can take an open neighbourhood to be one of the subbasis generators, $V(K,U)$. We need $U$ to be some open neighbourhood such that, each $s'\in U$ is, fibrewise, a three-plane that forms a graph over $s_x$. In other words, the plane corresponding to $s'(x)$ is just a small rotation of the initial three-plane, $s(x)$, for all $x\in Y$. We can take $K\subset Y$ to be any nonempty, compact subset, possibly $Y$ itself. The claim is that $p^{-1}(V(K,U))\cong V(K,U)\times\Maps(Y,SO(3))$. Indeed, if we fix a trivialisation over $s$, then by projection and Gram-schmidt, we are able to continuously assign an orthonormal trivialisation to each section in this neighbourhood and from there extract the isomorphism.  We therefore have a fibre bundle, $\scT\rightarrow\scC\rightarrow\scS$, or more explicitly
\begin{equation}\label{eq:AC3Sfibn}
    \Maps(Y,SO(3))\rightarrow \Gamma(Y,\cV_2(TY))\rightarrow \Gamma(Y,\cG(\varphi))\,.
\end{equation}

We do not know when, if ever, this space is in fact a trivial product, but we will show an example, \ref{sec:NCeg}, in which it is non-trivially fibred. More generally, it is natural to expect that $\Maps(Y,SO(3))$ has infinitely many components, because both $Y$ and $SO(3)$ have freely generated summands in third cohomology, while $G_2/SU(2)$ has only torsion, so it is plausible that the space of sections of $\cV_2(TY)$ also has only finitely many components. This leads us to expect that the fibre bundle is generically non-trivial, although this question is not settled here.

\section{Examples}\label{sec:examples}

In this section we will present example geometries that serve to illustrate concepts related with almost contact structures. The selected examples admit  $G_2$ structures with different properties, and some have been used in physics for the construction of supersymmetric solutions of string or M-theory.  We also explore the connection between ACM3S and associative submanifolds in a class of non-compact examples with $G_2$ holonomy.

\subsection{Heterotic $G_2$ systems on nilmanifolds}\label{sec:nilman}

Parallelizable nilmanifolds provide explicit examples of $G_2$ structure manifolds, that can be analysed in great detail using their left-invariant one-forms \cite{Fernandez:2008wla,delBarco:2020ddt}. Here we will recapitulate, in some detail, a particular example from Ref.~\cite{Fernandez:2008wla}: the nilmanifold $N(3,1)$, which solves the heterotic $G_2$ system presented in \ref{sec:het} and which appears particularily apt to illustrate that almost contact structures give added insight to the physical properties of string compactifications. 

Let $H(3,1)$ denote the seven dimensional generalised Heisenberg group., consisting of nilpotent real matrices of the form (cf. page 12 of \cite{Fernandez:2008wla})
\be
H(3,1) = \left \{
\left(\begin{array}{lllll}
1 & x_1 & x_2 & x_3 & z \\
0 & 1 & 0 & 0& y_1\\
0 & 0 & 1 & 0& y_2\\
0 & 0 & 0 & 1& y_3\\
0 & 0 & 0 & 0& 1\\
\end{array}
\right)
| x_i, y_i, z \in \mathbb{R}, 1<i<3
\right\}\,.
\ee
 From this, we may construct a compact nilmanifold $N(3,1)=\Gamma(3,1)\backslash H(3,1)$, where $\Gamma(3,1) \in H(3,1)$ consists of integer matrices of the above form. As all nilmanifolds, $N(3,1)$ is parallelizable, and its geometric features can be derived using a basis of left-invariant one-forms $e^a$, $a=1,..,7$ on $H(3,1)$
 which satisfy the structure equations
\be \label{eq:structeq}
\dd e^i = 0  \; , 1<i<6  \; \;  ,  \; \; \dd e^7 = a  e^{12} + b  e^{34} +c  e^{56}  \; ,
\ee
where we use the abbreviation $e^{ij} = e^i \w e^j$ and $a,b,c$ are real non-zero constants. Also, in this section, we will not distinguish between $\dd_7$ and $\dd$, since they coincide on $N(3,1)$. The structure equations clearly show that the $N(3,1)$ can be viewed as  a twisted torus  
\[
S^1 \hookrightarrow N(3,1) \rightarrow T^6 \; ,
\] 
and that $e^7$ acts as a connection that encodes the twisting of  $S^1$ over the six-torus $T^6$.

In Ref.~\cite{Fernandez:2008wla}, Lemma 5.5  shows that $N(3,1)$ admits a three-parameter family of  $G_2$ structures, defined by
\be \label{eq:nilphi}
\varphi = (e^{12}+e^{34}+e^{56})\w e^7 + e^{135}-e^{146}-e^{236}-e^{245}
\ee 
with associated diagonal metric $g_{\varphi}=e^1\otimes e^1+...+e^7\otimes e^7$. An explicit calculation shows that the Hodge dual fourform is
\[
\psi = *_{\varphi} \varphi = e^{3456} + e^{1256}+ e^{1234} - e^{2467}+e^{2357}+e^{1457}+e^{1367} \; .
\] 
Moreover,
\[ 
\dd \psi=0 \; , \; 
\dd \varphi = (a+b) e^{1234} + (a+c) e^{1256} + (b+c)e^{3456}
\]
and thus the torsion is contained in the classes $\tau_0$ and $\tau_3$. Explicitly, we have 
\be \label{eq:niltau0}
\tau_0 = \frac{2}{7} (a+b+c) \; , \text{and} \; \tau_3 = *\dd \varphi- \tau_0 \varphi \; .
\ee

Note that the case where $c=-(a+b)$ leads to a vanishing $\tau_0$ and hence a Minkowski solution to the heterotic $G_2$ system of section \ref{sec:het}. Furthermore, in this case, $T$ simplifies to
\be \label{eq:nilT}
T= -\tau_3= -(a+b) e^{567} +b e^{347} +a e^{127} \; .
\ee
 We will restrict to this case in the following. 

\subsubsection{Solving the heterotic $G_2$ system}
\label{sec:nilconnection}

In order to solve the heterotic $G_2$ system of section \ref{sec:het}, the nilmanifold $Y$ must admit $G_2$ instanton connections on $TY$ and $V$. 
Ref.~\cite{Fernandez:2008wla} shows that this can be accomplished on $N(3,1)$, provided that $c=-(a+b)$. This is seen as follows: recall that  all nilmanifolds have Levi-Civita connection 1-forms completely specified by their structure constants $f^i{}_{jk}$ :
\be
(\kappa^{LC})^i_j = \frac{1}{2}(f^i{}_{jk}-f^k{}_{ij}+f^j{}_{ki})e^k
\ee
which can be read off from \eqref{eq:structeq} using $\dd e^i = f^i{}_{jk} e^{jk}$.

 Moreover,   $\nabla^+ = \nabla^{LC} + \frac{1}{2} T$ is the unique $G_2$ compatible connection, and has  associated connection one-forms  \cite{Fernandez:2008wla} 

\be \label{eq:nilConn1form}
(\kappa^{+})^1_2 = - a e^7 \; \; , \; \;
(\kappa^{+})^3_4 = -b e^7 \; \; , \; \;
(\kappa^{+})^5_6 = (a+b)e^7 \; \; . \; \; 
\ee
As always, the curvature two-forms are given by $(\Omega)^i_j = \dd (\kappa)^i_j + (\kappa)^i_k \w  (\kappa)^k_j$, so  
\be
(\Omega^+)^1_2 = -a \dd e^7 
\; , \; (\Omega^+)^3_4 = -b \dd e^7
\; , \; (\Omega^+)^5_6 = -(a+b) \dd e^7 \; .
\ee

Now, as discussed in section \ref{sec:het}, supersymmetry requires that $\nabla^+$ is a $G_2$ instanton, and so  should satisfy \eqref{eq:Vinst}. 
This follows from 
\[
\dd e^7 \w \psi = (a+b-a-b) e^{123456} = 0 \; .
\]
  Thus, $a+b+c=0$ implies that $\nabla^+$  is a $G_2$ instanton. In fact, the same holds for any other connection with one-form components proportional to $e^7$.
  
In the same manner, we may then construct a vector bundle connection $A$ satisfying \eqref{eq:Vinst}. We assign connection one-forms  $(\kappa^{A})^i_j \sim e^7$, which will satisfy \eqref{eq:Vinst}. As shown in Ref.~\cite{Fernandez:2008wla}, the nilmanifold thus admits a 3-parameter family of $G_2$ instanton connections $A^{\lambda,\mu,\tau}$.
 Finally, with this choice of connections,  the Bianchi identity associated to the anomaly cancellation condition \eqref{eq:flux} 
 admits a solution as long as $(\lambda,\mu,\tau)\neq (0,0,0)$ and $\lambda^2+\mu^2+\tau^2 < a^2 + b^2 +c^2= 2(a^2+b^2+ab)$ \cite{Fernandez:2008wla}.

\subsubsection{Left-invariant contact structure and N=2 supersymmetry}

The left-invariant one-forms $e^i$ have dual vectors $E_i$ defined by
\[
 e^j(E_i) = \delta_i^j \; .
\] 
Having seven globally defined vectors is the hallmark of a parallelizable seven-dimensional manifold. Picking any of these vectors defines an ACMS on N(3,1), and any three-frame  defines an ACM3S. 

In this subsection, we will determine the properties of the ACMS defined by  
\[
R = E_7 \mbox{ and } \sigma = e^{7} \; .
\]
 By definition, we have $R(\sigma)=1$. Moreover, by \eqref{eq:structeq} we see that $\dd \sigma $ is purely transverse, so $\rm{Ker}(\sigma)$ is non-integrable and, therefore, not tangent to a 6-dimensional foliation (cf. section \ref{sec:ACStrans}).  Indeed, we have that
\be \label{eq:nilCS}
\sigma \w \dd \sigma  \w \dd \sigma \w \dd \sigma = -3! a \, b \, (a+b) \, e^{1234567} \neq 0  \; .
\ee
Thus, the almost contact structure induced by $\sigma$ is in fact a contact structure.   The fundamental two-form is 
\[
\omega= i_R(\varphi) = a e^{12} + b e^{34} - (a+b) e^{56} = \dd \sigma \; .
\]

This contact structure furthermore leads to enhanced supersymmetry. Recall, from section \ref{sec:N2}, that when $R$ is covariantly constant, there  exist two covariantly constant spinors on $Y$. We can phrase the covariant constancy of $R$ as a question on the connection one-forms. Indeed, the connection one-form is defined by 
\begin{equation} \label{eq:con1form}
    (\kappa^+)_{ji}(E_k):=g(\nabla^+_{E_k}E_j,E_i)
\end{equation}
and, consequentially, $\nabla^+ E_7=0$ if and only $(\kappa^+)_{7i}\equiv 0$.  That this is true follows immediately from the discussion leading to \eqref{eq:nilConn1form}.

We may go on to check the compatibility of the  $G_2$ instanton connections and $H$-flux, constructed in the previous subsection, with the possible existence of $N=2$ supersymmetry enhancement induced by the covariantly constant vector field, $E_7$. The connection one-forms  clearly lack a transverse piece, whereas the associated curvature is completely transverse, i.e.
\[
(\kappa^A)^i_j  \sim \sigma \; , \; (\Omega^A)^i_j \sim  a e^{12}+b e^{34}-(a+b) e^{56} \; .
\]
We thus read off that $F_0 = 0$, as we have argued in section \ref{sec:N2} should be the case for $N=2$ enhanced solutions. Likewise, the anomaly cancellation condition requires, for $N=2$ SUSY, that $H$ is completely transverse up to an exact contribution. Indeed, the torsion \eqref{eq:nilT} for this nilmanifold example lacks a transverse piece, and has 
\[
 T_0 = \dd \sigma = \dd \Sigma\; .
\]
We thus conclude that the $N(3,1)$ solution to the $N=1$ heterotic $G_2$ system, that was constructed in Ref.~\cite{Fernandez:2008wla}, in fact preserves $N=2$ supersymmetry. {We will discuss this further in the next subsection.}

\subsubsection{Left-invariant almost contact 3-structures}

In the preceding section we showed that $N(3,1)$ admits a left-invariant CS which is associated to two covariantly constant spinors, leading to $N=2$ supersymmetry. A natural question to pose is whether the ACS associated to the remaining left-invariant one-forms, $e^i$, for $i=1,..,6$, give a further enhancement of supersymmetry. To answer this question we are led to construct, and explore, the space of left-invariant ACM3S on this nilmanifold. As discussed in section \ref{sec:count-ac3s}, since we are dealing with a parallelisable manifold, we expect several distinguishable ACM3S, corresponding to the $7 \choose 2$ different ways of choosing 2-frames  among the seven left-invariant vectors $E_i$. 

As a first remark, we have $\dd e^i = 0$ for $i \neq 7$. Consequently, the ACS associated to $e^i$ for $i \neq 7$ are clearly not contact structures. Thus $N(3,1)$ does not admit a left-invariant  contact 3-structure (see the discussion at the end of \ref{ssec:ACS}).

Let us then start by identifying $R^1=E_7$. Then, we may choose $R^2=E_1$ say, after which $R^3$ will be determined by \eqref{eq:SU2-R3}  to be $R^3=E_2$. Choosing instead $R^2=E_3(E_5)$ gives $R^3=E_4(E_6)$, and if we start with picking $R^2=E_{2n}$ we find the same trivialisation up to an overall sign. The upshot is that, once $R^1=E_7$ is chosen, there are three inequivalent trivialisations  $(R^1,R^2,R^3)=(E_7,E_1,E_2)$,$(E_7,E_3,E_4)$,$(E_7,E_5,E_6)$. As discussed in section \ref{sec:count-ac3s}, these partially overlapping ACM3S are a consequence of the parallelizability of the nilmanifold. We will see below that these different trivializations all have the same qualitative properties.

For concreteness, let us study the trivialisation $(R^1,R^2,R^3)=(E_7,E_1,E_2)$. We may now determine whether the associated splitting $TY=\cT\oplus\cT^{\perp}$ lead to involutive $\cT, \cT^{\perp}$. As discussed in Section \ref{sec:su2int}, $\cT$ is involutive if for any one-form $\xi \in \cT^{*,\perp}$ we have
\[
i_{R^{\alpha}}i_{R^{\beta}} \dd \xi = 0 \; .
\]
This follows directly from the fact that $\cT^{*,\perp}=\text{Span}\{e^3,e^4,e^5,e^6\}$, and
\[
\dd e^i = 0 \; \;  \forall \; \; i \neq 7.
\]
In contrast, it is evident from the discussion around equation \eqref{eq:nilCS}, that $\cT^{\perp}$ fails to be involutive since 
\[
i_{R^1} i_{R^2} \dd \sigma^1 = a \neq 0 \; .
\]
This conclusion clearly holds also for the ACM3S given by
$(R^1,R^2,R^3)=(E_7,E_3,E_4)$ or $(E_7,E_5,E_6)$. We thus conclude that these ACM3S are associated three-dimensional foliations of $Y$, but no four-dimensional foliation. The leaves of the three-dimansional foliation are associative submanifolds (however, they are not volume-minimizing since we lack a calibrating three-form).

Let us now explore an ACM3S that does not include $E_7$ in the three-frame $(R^1,R^2,R^3)$. This leads to rather different properties. For example, let us take $(R^1,R^2,R^3)=(E_1,E_3,E_5)$. The associated splitting $TY=\cT \oplus \cT^{\perp}$ then lead to involutive $\cT$ and $\cT^{\perp}$. Clearly, since $\cT^{*,\perp}=\text{Span}\{e^2,e^4,e^6,e^7\}$, for any $\xi \in \Gamma(\cT^{*,\perp})$ we have either $\dd \xi = 0$ or
\[
i_{R^\alpha} i_{R^\beta} \dd \xi \sim i_{R^\alpha} i_{R^\beta} \dd e^7 = 
 0
\]
showing that $\cT$ is involutive. Moreover, $\cT^{\perp}$ is involutive, since  $(\sigma^1,\sigma^2,\sigma^3)=(e^1,e^3,e^5)$ are all closed. Thus, with this choice of ACM3S, we see that Y admits both a three- and a four-dimensional foliation. The leaves of these foliations are, respectively, associative and coassociative submanifolds. Furthermore, since $\dd\psi=0$, the leaves of the latter foliation correspond to calibrated submanifolds. 

Supersymmetry enhancement cares less about these different ACM3Ss on $N(3,1)$. Among the left-invariant basis vectors, only $E^7$ is covariantly closed. This follows directly from the definition of the connection one-form, \eqref{eq:con1form}, in conjunction with \eqref{eq:nilConn1form}. Thus, none of the left-invariant ACM3S under study  imply an enhanced supersymmetry to $N=4$. {Just like in section \ref{sec:N2}, where we saw that a reduction of structure group need not be compatible with a given connection, the parallelisability of the nilmanifold does not imply supersymmetry is maximally enhanced. Barring extra covariant vector fields on this nilmanifold,  the  holonomy group stays $SU(3)$. }

Finally, before closing this discussion, let us remark that we have not exhausted the possible almost contact structures. Our discussion is limited to left-invariant ACM3Ss, and even here one can imagine constructing ACM3S using  position dependent linear combination of the left-invariant forms which may lead to different conclusions.

\subsection{Barely $G_2$ examples}\label{sec:barelyG2}
In this section we demonstrate the features of an $SU(2)$ structure in a special class of $G_2$ structure manifolds known as barely $G_2$ manifolds,  \cite{joyce1996:2,Grigorian:2009nx, Harvey:1999as}. These manifolds are constructed from Calabi-Yau threefolds endowed with a real structure. For our purposes, we will focus on the subclass where the real structure is freely acting and where the initial Calabi-Yau manifold has vanishing Euler characteristic. Although this is expected to be an extremely small class (there are only two complete intersection Calabi-Yau manifolds with these properties, \cite{Grigorian:2009nx}, with Betti numbers (15,15) and (19,19)), the restriction is purely for convenience. We will begin reviewing the construction and key features of barely $G_2$ manifolds, before moving on to $SU(2)$ structures.

The starting point for a barely $G_2$ manifold is a Calabi-Yau threefold, $Z$, endowed with a real structure, that is, an antiholomorphic involution, $\zeta$. To avoid having to resolve singularities, we will assume that $\zeta$ is freely acting. This real structure is used to endow $Z\times S^1$ with a $\IZ_2$ action, $\hat{\zeta}=\zeta\times (-1)$, where $(-1)$ acts on $S^1\subset \IC$ via $e^{i\theta}\mapsto (e^{-i\theta})$, i.e. reflection about the real axis.  The barely $G_2$ manifold, $Y$ is the quotient by this action,
\begin{equation*}
  Y=(Z\times S^1)/(\zeta\times(-1))\,.
\end{equation*}

For us, it will be convenient to give an alternative definition for $Y$:
\begin{equation}
  Y=(Z\times [0,1])/\sim
\end{equation}
where we identify $(z,0)\sim (\zeta(z),1)$ and $\partial_t|_{(z,0)}\sim -\partial_t|_{\zeta(z),1}$.

Observe that this space has only a single constant spinor, but nevertheless the holonomy is a proper subgroup of $G_2$, being $SU(3)\rtimes\IZ_2$.  The induced $G_2$ structure forms are induced by the Calabi-Yau Kahler form, $\omega$, and holomorphic three-form $\Omega$:
\begin{align}
  \varphi&=\omega\w \dd t+\Rea\Omega\\
  \psi&=\rho-\dd t\w\Ima\Omega\,.
\end{align}
Observe that these product forms indeed survive the quotient, since:
\begin{align}
  \zeta^*\Omega&=\bar{\Omega}\\
  \zeta^*\omega&=-\omega\\
  (-1)^*\dd t&=-\dd t\,.
\end{align}

Now, by assumption, $Z$ has vanishing Euler characteristic and consequently the underlying smooth manifold admits a vector field, $v$. We will assume that we are given such a vector field, say $v$, and, without loss of generality, that it is of unit norm and invariant under the real structure, $\zeta^*v=v$.  This will guarantee that $v$ induces a vector field on the barely $G_2$ manifold, $Y$, say $R^1$.

  On the other hand, if $I$ denotes the complex structure on $Z$, then $w:=Iv$ is another unit vector field, but is not invariant under the real structure, instead acted on by a sign, $\zeta^*w=-w$. Similarly, the unit tangent vector field on the interval, $\partial_t$, is not invariant so is not well-defined on equivalence classes.

  The three vector fields, $v,w,\partial_t$, obviously induce vector fields on the product $Z\times [0,1]$, but as observed above, neither $w$ nor $\partial_t$ will survive the quotient individually. It is, however, possible to form invariant combinations using both vector fields,  in particular by rotating through the two-frame they span.  More precisely, consider
  \begin{equation}
    R^2=\cos(\pi t)w-\sin(\pi t)\partial_t\,,
  \end{equation}
  which is well-defined on the quotient since 
  \begin{equation*}
    R^2_{(z,0)}=w_z=R^2_{(\zeta(z),1)}\,.
  \end{equation*}
  Note, further, $R^1$ is everywhere orthogonal to $R^2$ so, in particular, it is everywhere linearly independent.

  The third vector field in the three-frame necessary to define an $SU(2)$ structure is obtained with the $G_2$ induced cross product:
  \begin{equation}
    R^3:=R^1\times_\varphi R^2
  \end{equation}
  though it is in fact easiest to compute first the dual one-form. Indeed, by definition $\sigma^3:=g(R^3,-)$ can be computed to be
  \begin{equation}
    \sigma^3=i_{R^2}i_{R^1}\varphi = \cos(\pi t)\dd t-\sin (\pi t)g(w,-)\,.
  \end{equation}
  This is now easy to dualise back to a vector field and we obtain
  \begin{equation}
    R^3 = -\sin (\pi t)w+\cos(\pi t)\partial_t\,.
  \end{equation}

  We have now fixed three linearly independent vector fields that fix the $SU(2)$ structure and can now directly compute the induced structure forms.   Before doing so, we remark that these vector fields can not be covariantly constant, since we have assumed that the initial Calabi-Yau manifold had full $SU(3)$ holonomy. As a consequence, the $G_2$ connection does not descend to a connection of the reduced $SU(2)$ structure.

  Each vector induces an $SU(3)$ structure, for which we can construct a two-form $\omega_i:=i_{R^i}\varphi$ and three-form $\Omega_-^{i}=i_{R^i}\psi$, a dual one-form $\sigma_i=g(R^i,-)$ and endomorphism $J^i$ such that $g(J^i-,-)=\omega_i$.  We will focus on computing $\omega_i$ and $\Omega_-^i$ since we do not need a metric for this data.  The computations are straightforward and we obtain:
  \begin{align}
    \omega_1&=-w^\#\w \dd t+i_v\Rea\Omega\\
    \omega_2&=-\sin(\pi t)\omega+\cos(\pi t)(v^\#\w \dd t+i_v\Ima\Omega)\\
     \omega_3&=\cos(\pi t)\omega-\sin(\pi t)(v^\#\w \dd t+\Ima\Omega)\\
     \Omega_-^1&=-w^\#\w \omega+\dd t\w i_v\Ima\Omega\\
    \Omega_-^2&=\sin(\pi t)\Ima\Omega+\cos(\pi t)(v^\#\w\omega+\dd t\w i_v\Rea\Omega)\\
    \Omega_-^3&=-\cos(\pi t)\Ima\Omega-\sin(\pi t)(v^\#\w \omega+\dd t\w i_v\Rea\Omega)
  \end{align}
  where we have introduced $v^\#=g(v,-)$ and $w^\#=g(w,-)$ and recall that $\omega,\Omega$ are the Calabi-Yau structure forms.  The particular vector field that we give here can be viewed as rotating through the local frame $(v,w,\partial_t)$, a fact that is also apparent in the structure forms.  On a generic barely $G_2$-manifold we would not expect to find vector fields analogous to $v$ and $w$, but even so we are guaranteed to find a ACM3S. One would expect that any such three-framing would intertwine the Calabi-Yau and interval directions much more intricately than exhibited here.

  We can now consider the space of trivialisations that are compatible with the splitting induced by $(R^1,R^2,R^3)$. We will show that there are at least two connected components, which we distinguish by map induced on the first homology groups. As is well-known, the first homology group of $SO(3)$ is $H_1(SO(3),\IZ)\cong\IZ_2$, with generator induced by the generator of $SO(2)\cong S^1$ under the inclusion $SO(2)\hookrightarrow SO(3)$, i.e.
  \begin{align}
      \rho:I\rightarrow SO(3)\,,&&\rho(t)=\left(\begin{array}{ccc}1&0&0\\ 0&\cos (2\pi t) &-\sin(2\pi t)\\ 0&\sin(2\pi t)&\cos 2\pi t
      \end{array}\right).
  \end{align}
  Using standard techniques in algebraic topology, it can be shown that $H_1(Y,\IZ)\cong\IZ_2$ also, with generator induced by the open path on $Z\times S^1$
  \begin{equation}
      \chi(t)=\left\{\begin{array}{cc}
          (\gamma_0(2t),1) &\quad t\in[0,1/2]  \\
           (\zeta(x_0),e^{i\pi(2t-1)})& \quad t\in[1/2,1]
      \end{array}\right.
  \end{equation}
  with $\gamma_0$ an arbitrary path in $Z$ between a fixed $x_0$ and $\zeta(x_0)$. Note that since $Z$ is simply connected, the homotopy class of $\chi$ is independent of $\gamma_0$.
  
  Consider, then, the map $\Theta:Y\rightarrow SO(3)$, induced by the map, $\tilde{\Theta}$ on $Z\times I$ 
  \begin{equation}
      \tilde{\Theta }(x,t)=\left(\begin{array}{ccc}1&0&0\\ 0&\cos (2\pi t) &-\sin(2\pi t)\\ 0&\sin(2\pi t)&\cos 2\pi t
      \end{array}\right).
  \end{equation}
  Then, it is straightforward to check that
  \begin{equation*}
      \Theta_*\chi(t)=\left\{ \begin{array}{cc}1&\quad t\in[0,1/2]\\
      \Theta(\sigma(x_0),2t-1)&\quad t\in[1/2,1]\end{array}\right.\sim \rho(t).
  \end{equation*}
  Consequentially, $\Theta_*:H_1(Y;\IZ)\rightarrow H_1(SO(3),\IZ)$ is an isomorphism and $\Theta$ can not be homotopic to the constant map $1:Y\rightarrow SO(3)$. The trivialisation corresponding to $\Theta$ is given by the vector fields $(S^1,S^2,S^3)$, which, by direct computation, are:
  \begin{align}
      S^1 &= R^1=v\\
      S^2&=\cos(2\pi t)R^2-\sin(2\pi t) R^3=\cos (\pi t)w-\sin (3\pi t)\partial_t\\
      S^3&=\sin(2\pi t)R^2+\cos(2\pi t)R^3=-\sin (\pi t)w + \cos(3\pi t)\partial_t
  \end{align}
  The fact that $\Theta$ does not induce the zero map on first homology, or equivalently, that it does not induce the zero map on the fundamental group, implies that $\Theta$ can not be lifted to the universal cover of $SO(3)$, which is $SU(2)$.  This indicates that the spin structure that is canonically associated to the trivialised bundle, $\cT$, differs between the $R$ and $S$ trivialisations. 
  
  Although $\Theta$ induces countably many trivialisations via $\Theta^n,\,n\,\in\,\IZ$, the first homology groups are only able to distinguish between the cases that $n$ is even or odd. The group $SO(3)$ has only torsion homotopy groups in dimensions zero through seven, with the exception of $\pi_3$, $\pi_3(SO(3))\cong\IZ$. Therefore, the third homotopy group is a natural arena in which to attempt to distinguish the components of these maps, though we do not attempt that here.
  
  Further, we have only shown that these two trivialisations are non-homotopic inside the fixed trivial 3-bundle and it does not necessarily follow that they are non-homotopic in the space of all ACM3S's. This would only follow if the locally trivial product structure that we identified in Subsection \ref{sec:count-ac3s} is in fact trivial. Working with the space of all ACM3S's is more subtle because they correspond to sections of a possibly non-trivial bundle, with twisting dictated by the initial $G_2$ structure and we have nothing conclusive to say about this.
  
\subsection{A compact $G_2$ holonomy example}
{Let us now turn to manifolds with $G_2$ holonomy, and explore almost contact structures on such spaces. In this section, we review an example of compact seven-manifolds with $G_2$ holonomy that is due to Joyce.  As we will see, this construction allows us to specify a canonical AC(3)S that is suitable to illustrate the topological nature of the space of ACM3S. In the next section, we will explore this topic in more detail in a non-compact setting.}

The construction of this type of $G_2$ holonomy manifolds is modelled on the Kummer construction of $SU(2)$ holonomy metrics on the orbifold $T^4/\mathbb{Z}^2$. The construction was originally proposed in \cite{joyce1996:1,joyce1996:2}, has been presented in great detail in the books \cite{joyce2000}, and a clear summary of the construction can be found in Ref.~\cite{joyce:proc}. For the sake of completeness, we will recapitulate the construction algorithm here.

To begin, we start with the 7-dimensional torus $T^7$, with coordinates $x^a, a=1,..7$ satisfying $x^a \sim x^a+1$. This space may be equipped with the standard $G_2$ structure $\varphi_0$ as in \eqref{eq:g2standard}, by the global identication $e^a = \dd x^a$. The metric $g_{\varphi_0}$ is naturally flat. By quotienting $T^7$ by a finite automorphism group $\Gamma$ that respects $\varphi_0$, a new $G_2$ space is obtained. In general, $T^7/\Gamma$ will be an orbifold, with singular set $S$ specified by the fix points of $\Gamma$. We will see an example of this momentarily. 

In order to obtain a smooth $G_2$ manifold, there is then need to resolve the singularities. In Joyce's construction, this is accomplished by noticing that for certain groups $\Gamma$ the singular set $S$ decomposes into connected components that may be resolved using standard procedures in complex algebraic geometry (see below). This resolution gives a  non-singular 7-manifold $Y$, which can be shown to admit a 1-parameter family of $G_2$ structures $\varphi_t$ with torsion $|\nabla \varphi_t|= {\mathcal O}(t^4)$. This 1-parameter family of $G_2$ structures is obtained by gluing together, using a partition of unity, the flat $G_2$ structure $(\varphi_0,g_0)$ in the non-singular ``bulk'' of $Y$ with local $G_2$ structures $(\varphi_i,g_i)$ valid near the various resolved orbifold singularities. The non-zero torsion is localized in the regions with non-trivial derivatives for the partition of unity, i.e. where the resolved singular spaces adjoin with the bulk.

It is then possible to prove that for all sufficiently small parameters $t$, one may deform $(\varphi_t,g_t)$ to $(\tilde{\varphi},\tilde{g})$ with vanishing torsion. Finally, given the specific choices made in the construction, one may show that the holonomy of $\tilde{g}$ is indeed $G_2$, and not a subgroup thereof. Thus $Y$ is a compact $G_2$ holonomy manifold. 

Our purpose in this section is to explore AC(3)S that are compatible with Joyce's construction. Let us therefore describe in some more detail the singular set $S$, following \cite{joyce:proc} (see also \cite{Kronheimer:1989zs,kronheimer1989,ROAN1996489}). By careful selection of $\Gamma$, one may ascertain that $S$ decomposes into connected components $S_i$ that are locally isomorphic to either $T^3 \times \mathbb{C}^2/G$, for $G \subset SU(2)$ finite, or $S^1 \times \mathbb{C}^3/G$, for $G \subset SU(3)$ finite and freely acting on $\mathbb{C}^3 \backslash 0$. We may then use that 
\begin{enumerate}
\item \label{item:ale} $\mathbb{C}^2/G$, for $G \subset SU(2)$ finite, may be resolved to a smooth Asymptotically Locally Euclidean (ALE) space $U_2$, with K\"ahler metric of $SU(2)$ holonomy
\item $\mathbb{C}^3/G$, for $G \subset SU(3)$ as above, may be resolved to a smooth ALE space $U_3$, with K\"ahler metric of $SU(3)$ holonomy
\end{enumerate}
Then, we may resolve $Y$ by locally excising the connected components $S^\alpha$ of the singular set $S$, and then glue in smooth product spaces $\hat{S}^\alpha$, which have the form  $T^3 \times U_2$ or $S^1 \times U_3$, depending on the nature of the singularity. 

The product spaces $\hat{S}^\alpha$ admit $G_2$ structures, which we will discuss in detail for the example below. An obvious effect of this construction is that these local $G_2$ structures are reduced to $SU(2)$ and $SU(3)$ structures, respectively, and the holonomy of the local metric will be either $SU(2)$ or $SU(3)$. This provides a first link to the AC(3)S that we have discussed in this paper. Let's explore that in more detail in an example.

The following example is taken from \cite{joyce1996:1}: as above, we construct an orbifold by quotienting $T^7$ by a finite group, which in this example is the $\mathbb{Z}_2^3$ generated by\footnote{There is a change in convention in the definition of the local form of $\varphi$ with respect to \cite{joyce1996:1}; this requires changing a sign in the $\gamma$ action.} 
\begin{align}
    \alpha((x_1,...,x_7)) &= (-x_1,-x_2,-x_3,-x_4,x_5,x_6,x_7) \\
    \beta((x_1,...,x_7)) &= (-x_1,\frac{1}{2}-x_2,x_3,x_4,-x_5,-x_6,x_7) \\
    \gamma((x_1,...,x_7)) &= (\frac{1}{2}-x_1,x_2,\frac{1}{2}-x_3,x_4,x_5,-x_6,-x_7) \; .
\end{align}
One can readily show that this group preserves the $G_2$ threeform $\varphi_0$ given in \eqref{eq:g2standard}. The singular set $S$ of $T^7/\mathbb{Z}^3_2$ is determined by the fix points of the generators $\alpha, \beta$ and $\gamma$, and a little thought reveals that $S$ decomposes into 12 disjoint components of the form $T^3 \times \mathbb{C}^2/\{\pm 1\}$ \cite{joyce1996:1}.

Now, we're in the first situation described in the above list, and for each simply connected component of singular set, we may blow up $\mathbb{C}^2/\{\pm 1\}$ to $U^A$, where $U^A$ is a smooth ALE space that agrees with $\mathbb{C}^2/\{\pm 1\}$ on its boundary. We may construct this geometry explicitly. Denoting by $(z^1, z^2)$ the coordinates on $\mathbb{C}^2$, $U^A$ admits a hyper-Kähler triple of two-forms $\{\omega^{\alpha}(t)\}$ satisfying
\be \label{eq:omegaiEH}
\omega^1(t) = \frac{i}{2} \partial \bar{\partial} f_t \; , \; 
\omega^2(t) + i \omega^3(t) = \dd z^1 \w \dd z^2
\ee
where we have introduced a suitable K\"ahler potential that reduces to $f_t\to |z_1|^2+|z_2|^2$ at the boundary of $U^A$, and hence guarantees that  the two-forms $\omega^{\alpha}(t)$ become flat $\hat{\omega}^{\alpha}$ at this boundary.

We then find that the smooth manifold $Y$, that result from desingularising $T^7/\mathbb{Z}^2_3$ in the manner just described, admit a nowhere vanishing three-form $\varphi_t$. Using the above defined two-forms, and taking one-forms $\sigma^{\alpha}$ as sections of $\Lambda^*T^3$, this can be written as
\[
 {\varphi}_t={\varphi}_0=\frac{1}{3!} \epsilon_{\alpha \beta \gamma}  \sigma^{\alpha} \w \sigma^{\beta} \w \sigma^{\gamma}
+ \sum_{\alpha}\sigma^{\alpha} \wedge \hat{\omega}^{\alpha} \; ,
 \]
 in the non-singular bulk of $T^7/\mathbb{Z}^2_3$, and
\[
 {\varphi}_t=\frac{1}{3!} \epsilon_{\alpha \beta \gamma}  \sigma^{\alpha} \w \sigma^{\beta} \w \sigma^{\gamma}
+ \sum_{\alpha}\sigma^{\alpha} \wedge \omega^{\alpha}(t) \; ,
 \]
 in an open neighbourhood that contains $S \times U^A$. This is a smooth three-form, since $\{\omega^{\alpha}(t)\} \to \{\hat{\omega}^{\alpha}\}$ at the boundary of $U^A$, and it is closed for all $t$. However, a subtlety of the construction is that it is only for small $t$ that $\varphi_t$ is guaranteed to be a positive three-form, and hence defines a $G_2$ structure, in the interpolating region \cite{joyce1996:1}. Finally, while $\varphi_t$ is closed for all $t$, it fails to be coclosed in this region. As discussed above, a closed and coclosed $G_2$ structure may be constructed, in the $t \to 0$ limit, as $\varphi = \varphi_t + \dd \eta_t$, where the two-form $\eta$ satisfies a certain elliptic differential equation that we will not discuss in detail.
 
Thus, in the parametric limit of small $t$, we have constructed a $G_2$ structure $\varphi_t$ which is of the form \eqref{eq:su2decomp-phi}, and provide an example of an ACM3S. In particular, there is a set of globally defined two-forms, $\{\omega^\alpha(t)\}$, which, when combined with the $G_2$ three-form, uniquely defines three one-forms $\{\sigma^\alpha\}$, which in turn defines an orthonormal three-frame $\{R^1,R^2,R^3\}$. 

So far, we have seen that there are obvious similarities between Joyce's torsionful $G_2$ structure $\varphi_t$ and the ACM3S decomposition that we know should exist for any  $G_2$ structure.  A more interesting question to ask is whether Joyce's construction automatically provides information about the AC(3)S of the torsion-free $G_2$ structure $\varphi_t$. This is true, but only to some extent. Indeed, any nowhere vanishing vector field $R$ will provide an ACS when combined either with $\varphi_t$ or $\varphi$. However the fundamental two-forms given by \eqref{eq:fundform} depend on the three-form, and hence the transverse geometry will differ between these ACS. Clearly, comparing to \eqref{eq:Wtau}, the $SU(3)$ torsion classes will also depend on whether the torsionful or torsionfree $G_2$ structure is selected.

For AC3S the situation is similar, but here one also has to take into account that  while linear independence of vector fields does not depend on the $G_2$ structure, orthonormality of vectors  do. Thus, an orthonormal three-frame $\{R^1,R^2,R^3\}$ associated to $\varphi_t$ will not be orthonormal with respect to $\varphi$. However, the space of AC3S, $\scC$, is topological, and hence the same for $\varphi_t$ and $\varphi$. Thus, we may hope to determine this space using only the explicit, torsionful $G_2$ structure $\varphi_t$. This is interesting, because we can then hope to even say something new about how to count associatives on a $G_2$ manifold. We hope to return to this intriguing problem in a future publication.

\subsection{A class of non-compact $G_2$ holonomy examples}\label{sec:NCeg}

In this section we will consider the local neighbourhood of an associative three-cycle in the context of ACM3Ss. Associative cycles are known to be relevant to M-theory compactifications and we earlier found a surprising relation with ACM3Ss.  In particular, the associated trivial bundle, $\cT$, can only be tangent to a submanifold if it is an associative submanifold.\footnote{Note that $\cT$ may be tangent to a submanifold without everywhere being a foliation, that is, the involutivity condition \eqref{eq:Tinvol} may be satisfied for $x\in X^3$, but not for arbitrary $x\in Y$.}  The question naturally arises: if we fix an arbitrary associative three-cycle, can we always choose a global ACM3S that restricts to the tangent bundle of this three-cycle? In the simplest case, one would be able to locally deform a given ACM3S in a neighbourhood of the chosen three-cycle to satisfy this condition and we explore this possibility using our explicit expression for the space of ACM3Ss. 

First, we will show by construction that there is always an ACM3S, defined in the neighbourhood of an associative three-cycle, which restricts to a trivialisation of the three-cycle's tangent bundle. This example, however, has boundary behaviour that is manifestly dependent on the chosen trivialisation at $X$. This is undesirable from the perspective of the initial compact manifold. We will therefore fix boundary conditions and study the space of compatible ACM3S. We will see that there are topologically distinct ACM3Ss at the boundary and, therefore, the possible configurations over the associative three-cycle depends on what happens at the edge of its local neighbourhood. Note that the question of which boundary conditions may arise depends on the global setting. It is an interesting open problem to determing what global topological features may obstruct choosing the ACM3S to be tangent to $X$. Finally, we will then study the space of all ACM3Ss, $\scC(NX,\varphi)$ and show, in particular, that the bundle structure observed in subsection \ref{sec:count-ac3s}, is non-trivial. Finally, we will fix boundary conditions on $NX$ and study the corresponding space of ACM3S, $\scC_\partial(NX)$, say. 

\subsubsection{Constructing an ACM3S}
Let $(Y,\tilde{\varphi})$ be a manifold with $G_2$ holonomy, i.e. $\varphi$ is a closed and coclosed stable three-form. Let $X\hookrightarrow \tilde{Y}$ a smooth associative three-cycle, meaning that the induced volume form on $X$ is precisely the restriction of the stable three-form: $\tilde{\varphi}|_X={\rm Vol}_X$. $X$ has a normal bundle, $NX$, and by standard results, this bundle is diffeomorphic to a tubular neighbourhood of $X$ in $\tilde{Y}$. By choosing such a diffeomorphism and pulling back $\tilde{\varphi}$, the normal bundle is endowed with a $G_2$ structure.

In practice, it will be most useful for us to replace $NX$ with a finite-radius disc bundle. This is for technical convenience and we will tend not to keep explicit track of the difference.  For concreteness, we will choose the embedding of $NX\hookrightarrow \tilde{Y}$ to be given, fibrewise, by geodesics.  That is, for $n_x\in NX,$ a small enough normal vector at $x\in X$, the corresponding point in the embedded submanifold is given by $\exp_x(n_x)\in Y$, where $\exp$ is exponential map of Riemannian geometry.

Pulling back the three-form, $\tilde{\varphi}$ we obtain a three-form on $NX$, $\varphi$, which is covariantly constant with respect to the Levi-Civita connection.  This has the convenient consequence that the value of $\varphi$ everywhere on $NX$ is fixed by its value on $X$, since we can parallel transport along the fibres:
\be 
\varphi_{(n,x)}={\sf P}_{\exp_x(tn)}(\varphi),
\ee
where ${\sf P}_\gamma$ indicates parallel transport along $\gamma$; our choice of path, $t\mapsto \exp_x(tn)$ is not unique, but $\varphi$ is completely independent of this choice.

We can now look at ACM3Ss compatible with the $G_2$ structure we have introduced. We are particularly interested in ACM3Ss that restrict to the tangent bundle of $X$, so our first task is to ensure that we can extend such a choice over the whole $NX$. 

In this example, there is a canonical choice for extending a trivialisation of $X$. Let us fix an oriented, orthonormal framing of $X$, thereby inducing a three-frame on $TY|_X$, say $(R^1_X,R^2_X,R^3_X)$. We need to extend this  three-frame over $NX$, whilst preserving $R^1_X \times_\varphi R^2_X=R^3_X$.  The obvious thing to do is parallel transport along the geodesics:
\begin{equation}
  R^i_{(x,n_x)}={\sf P}_{\exp_x(n_x)}R^i_{X,x}\,.
\end{equation}

This preserves \eqref{eq:SU2-R3}, since $\varphi$ is also given by parallel transporting along the geodesics off of $X$. On the other hand, the triple $(R^1,R^2,R^3)$ need no longer be integrable. Indeed, a straightforward computation shows that the Lie bracket of parallel transported vectors is given by:
\begin{equation}
    [{\sf P}R,{\sf P}S]=(\nabla_{{\sf P}R}{\sf P})(S)-(\nabla_{{\sf P} S}{\sf P})(R)+{\sf P}(\nabla_{{\sf P} R}S-\nabla_{{\sf P}S}R)\,,
\end{equation}
showing that parallel transport does not play well with the Lie bracket. This is not at all surprising: if integrability were preserved along the normal bundles, it would indicate a family of deformations of an associative three-cycle and finding such deformations is notoriously difficult, \cite{mclean1998deformations, Joyce:2016fij}. Integrability is not a requirement for us, however, and we are satisfied with an implicit, canonical ACM3S for each framing of the calibrated cycle $X$.

\subsubsection{Compatibility of boundary conditions}

We note that the ACM3S constructed in the previous section has a non-trivial relationship between the behaviour at the boundary of $NX$ and the precise framing of $X$. If we want to glue this back into the original, compact manifold, this is not satisfactory. We therefore consider a similar problem, but with fixed boundary conditions. Our approach will use the topological features of $\scC$ as opposed to the direct construction used above. 

To this end, let us fix an ACM3S at the boundary of $NX$, say ${\bf R}_\partial\in \scC(\partial(NX))$, where $\scC(\partial(NX)):=\Gamma\big(\partial(NX),\cV_2(T(NX))|_{\partial (NX)}\big)$, by abuse of notation. We will also fix boundary conditions at the zero section ${\bf R}_X\in \scC(X)$. Note that we have not actually imposed that $R^i_X$ come from the tangent bundle over $X$, despite our motivation for studying this problem.  Our conclusions are independent of which boundary conditions we impose at either end: it matters only that not all conditions are mutually compatible. 

To impose the boundary conditions over $X$, consider the manifold $Y^\circ$, which is obtained from $NX$ by cutting out the zero section. It turns out that $Y^\circ$ is very simple. Indeed, the normal bundle, $NX$ can be seen as a vector bundle associated to an $SU(2)\times SU(2)$-principle bundle over $X$. Since $SU(2)$ is simply connected and $X$ is three-dimensional, any such bundle is trivial and $NX$ is diffeomorphic to the product $X\times D^4$. Consequentially, we can identify $Y^\circ$ with $X\times S^3\times I$ for an open interval, $I$. 

We recall from Subsection \ref{sec:count-ac3s} that the space of ACM3Ss is the space of sections of a fibre bundle with typical fibre the homogeneous space $V_2(\IR^7)=G_2/SU(2)$. In our case, the seven manifold is parallelisable so that the fibre bundle is trivial and we can view our boundary conditions as maps ${\bf R}_\partial:X\times S^3\times \{1\}\rightarrow V_2(\IR^7)$ and ${\bf R}_X:X\times S^3\times \{0\}\rightarrow V_2(\IR^7)$. Of course, for consistency, it must be that ${\bf R}_X$ be constant in the fibral $S^3$ direction. An ACM3S that extends these boundary conditions is the same as a map $Y^\circ \rightarrow V_2(\IR^7)$ that restricts to ${\bf R}_\partial$ and ${\bf R}_X$. To put it another way, such a map is a homotopy of maps $X\times S^3\rightarrow V_2(\IR^7)$, so our problem is to determine the connected components of the space $\Maps(X\times S^3,V_2(\IR^7))$. We will utilise elementary topology techniques to put very coarse bounds on the number of these components. Readers unfamiliar with these techniques can consult e.g. \cite{hatcher2000algebraic} for a comprehensive introduction.  

To make things more concrete, we will assume our associative three-cycle is the three-sphere $X=S^3$. Since this is the simplest compact three-manifold, we would expect that any obstructions appearing in this example would occur for a generic associative submanifold. The first observation we make is that $V_2(\IR^7)$ is simply connected, which we shall show below using sequence chasing in a homotopy exact sequence. Given this fact, we know that the homotopy classes of maps $\Maps(X\times S^3,V_2(\IR^7))$ (i.e. the connected components of this space) is identical with the homotopy classes of basepoint preserving maps $\Maps_*(X\times S^3,V_2(\IR^7))$, so we will regard both spaces as coming with an arbitrary choice of basepoint.

Since maps out of a product space are not very convenient to deal with, we consider the fibration sequence
\begin{equation}
    S^5\rightarrow S^3\vee S^3 \rightarrow S^3\times S^3\rightarrow S^6\rightarrow \Sigma(S^3\vee S^3)\,. \label{eq:spherefibn}
\end{equation}
We have utilised several basic topological constructions here, which we briefly recall. In particular, the operations of wedge sum, $\vee$, the smash product, $\wedge$, and suspension, $\Sigma$, have appeared. Briefly, the wedge sum of pointed spaces, $(X,x_0),(Y,y_0)$ is defined to be quotient space $X\vee Y = (X\sqcup Y)/(x_0\sim y_0)$. In other words, the wedge sum is the space formed by gluing the two spaces together at the basepoint. The wedge sum of two circles is, for instance, the figure-eight space. Next, the smash product of two spaces is formed by quotienting the wedge sum out of the cartesian product, $X\w Y =(X\times Y)/(X\vee Y)$. This uses the embedding of the wedge sum $X\vee Y\hookrightarrow X\times Y$, which is induced by the maps $X\mapsto X\times \{y_0\}$ and $Y\mapsto \{x_0\}\times Y$. In the example of two circles, this embeds the wedge sum into the two-torus $T^2$ as the union of the longitudinal and meridianal circles, and the smash product is given by crushing these circles to a point. One can check that the resulting space is a homeomorphic to the 2-sphere.  Finally, the suspension of a space can be given by smashing with a circle, $\Sigma X:=S^1\w X$. The example with two circles exhibits the general property that the suspension of an $n$-sphere is homeomorphic to an $(n+1)$-sphere, $\Sigma(S^n)\cong S^1\w S^n\cong S^{n+1}$.  

Returning to the sequence, \eqref{eq:spherefibn}, we can now define the maps. In particular, the excerpt $S^3\vee S^3\rightarrow S^3\times S^3\rightarrow S^6$ is precisely the defining maps of the smash product: the first map is the canonical inclusion of the wedge sum into a product, and the final map is the quotient $S^3\times S^3\rightarrow S^3\w S^3\cong S^6$.  The first map can be seen as the boundary of the attaching map $\Phi: D^6\rightarrow S^3\vee S^3$ as part of CW construction for $S^6$ (see, e.g. \cite[p.175]{felix2012rational} for more details on this). Finally, the map $S^6\rightarrow \Sigma(S^3\vee S^3)$ is the suspension of this attaching map, $\partial\Phi$. It is, in particular, nullhomotopic, which gives us an excerpt of an exact sequence:
\begin{equation}
\begin{split}
      0\rightarrow \pi_0\big(\Maps_*(S^6,V_2(\IR^7)\big)\rightarrow\pi_0\big(\Maps_*&(S^3\times S^3,V_2(\IR^7))\big)\\
      &\rightarrow \pi_0\big(\Maps_*(S^3\vee S^3,V_2(\IR^7))\big)\label{eq:V2exact}
\end{split}      
\end{equation}
where the third term is the set we are interested in.  The first term is the homotopy group $\pi_6(V_2(\IR^7))$ and the last one is the direct sum $\pi_3(V_2(\IR^7))^{\oplus 2}$.  These homotopy groups can be calculated using the fibration $SU(2)\rightarrow G_2\rightarrow V_2(\IR^7)$ and the associated long exact homotopy sequence, as will see now. 

Let us first prove the earlier claim that $V_2(\IR^7)$ is simply connected. Indeed, as an excerpt from the exact sequence we have:
\begin{equation}
    \pi_1(G_2)\rightarrow \pi_1(V_2(\IR^7))\rightarrow\pi_0(SU(2))\,,
\end{equation}
and the fact that both $G_2$ and $SU(2)$ are simply connected shows that $\pi_1(V_2(\IR^7))$ is indeed vanishing.

Let us now look for the groups appearing as a consequence of \eqref{eq:spherefibn}, i.e. $\pi_3(V_2)$ and $\pi_6(V_2)$. First, we use the following extract:
\begin{equation}
  0\rightarrow \pi_7(V_2)\rightarrow \pi_6(SU(2))\rightarrow \pi_6(G_2)\rightarrow \pi_6(V_2(\IR^7))\rightarrow \pi_5(SU(2))\rightarrow 0
\end{equation}
where the fact that $\pi_5(G_2)=0=\pi_7(G_2)$, \cite{mimura1967homotopy} has been used. Since $\pi_5(SU(2))=\IZ_2$, exactness implies that $\pi_6(V_2(\IR^7))\neq 0$. Further, $\pi_6(G_2)=\IZ_3$ and $\pi_6(SU(2))=\IZ_{12}$ so both $\pi_6(V_2(\IR^7))$ and $\pi_7(V_2(\IR^7))$ are certainly torsion, in particular having finite cardinality.

Similarly,  we can extract the exact sequence
\begin{equation}
    0\rightarrow \pi_4(V_2)\rightarrow \pi_3(SU(2))\rightarrow \pi_3(G_2)\rightarrow \pi_3(V_2)\rightarrow 0
\end{equation}
where $\pi_3(G_2)\cong \pi_3(SU(2))\cong\IZ$. A map from $\IZ\rightarrow \IZ$ is either zero, or injective. In the first case, we would find $\pi_4(V_2)\cong\pi_3(V_2)\cong\IZ$, while in the second we would conclude that $\pi_4(V_2)=0$ and $\pi_3(V_2)$ is torsion. In fact, this map can not be zero. This is because $SU(2)\cong S^3$, so if the generator of $\pi_3(SU(2))$ maps to zero in $\pi_3(G_2)$ we have to conclude that the image of $SU(2)$ inside of $G_2$ is nullhomotopic. In particular, all maps $\pi_i(SU(2))\rightarrow\pi_i(G_2)$ would have to vanish, which is known not to happen, \cite{mimura1967homotopy}.  Therefore, $\pi_4(V_2)=0$ and $\pi_3(V_2)$ is torsion. 

With these results in hand, we can return to \eqref{eq:V2exact}. Since $\pi_6(V_2)\neq 0$ is injected into $\pi_0(\Maps_*(S^3\times S^3))$, it follows that this set must have cardinality at least as large as $\pi_6(V_2)$.  Further, exactness implies this map surjects onto a subgroup of $\pi_3(V_2)^{\oplus 2}$ and we have just seen that this is torsion. As a consequence, the set of connected components, $\pi_0(\Maps_*(S^3\times S^3,V_2(\IR^7)))$, is non-empty but finite.

This implies that there are indeed topological obstructions preventing us from extending arbitrary boundary conditions, but there are only finitely many distinct components, or at least only finitely many that are detected by the induced maps on homotopy groups.  

Assuming that we have found compatible boundary conditions, we can still ask about the space of ACM3S extending them. Our above argument shows that this is the space of paths in $\Maps(S^3\times S^3,V_2(\IR^7))$, with fixed endpoints.  For simplicity, let us assume that our boundary conditions are such that ${\bf R}_X={\bf R}_\partial\sim\,{\rm constant}$.  Such paths are the same as maps on the cylinder $S^3\times S^3\times I$, constant at the interval endpoints. In fact, we can think of this as a map on the suspension $\Sigma(S^3\times S^3) \cong \Sigma(S^3)\vee \Sigma(S^3)\vee \Sigma(S^3\wedge S^3)$. Therefore, the distinct components of this space are 
\begin{equation}
    \pi_0(\Maps(\Sigma(S^3\times S^3),V_2(\IR^7)))\cong\pi_4(V_2(\IR^7))^{\oplus 2}\oplus \pi_7(V_2(\IR^7))=\pi_7(V_2(\IR^7))\,.
\end{equation}
We have used the earlier observation that $\pi_4(V_2)=0$. The fact that $\pi_7(V_2)$ is torsion means that there are finitely many distinct classes of ACM3Ss with these boundary conditions. 

In summary, we have seen that boundary conditions on $NX$ may obstruct our ability to find an ACM3S that is tangent to the associative three-cycle, $X$. It is an interesting question for future work to determine these obstructions more precisely and to see how the global topology of $\tilde{Y}$ influences the ACM3S that are obtainable on the boundary $NX$.

\subsubsection{All ACM3S's on $NX$}
We will now return to study the space of all ACM3Ss on $NX$, without imposing any boundary conditions.  This is interesting to study because the topological simplicity of $NX$ means we can easily see that the fibre bundle structure on the space of ACM3S's cannot  be trivial.

We will continue to assume that $X$ is a three-sphere to keep the calculations as simple as possible.  In this case $NX\cong X\times \IR^4$, and the space of all ACM3Ss is given by the space of maps into $V_2(\IR^7)$, $\scC(NX)=\Maps(NX,V_2(\IR^7))$. The connected components of this space are the homotopy classes of these maps and contractibility of $\IR^4$ means that this is the same as the homotopy classes of maps $\Maps(X,V_2(\IR^7))$. Conveniently, this is the homotopy group $\pi_3(V_2(\IR^7))$, which was argued above to be finite.  

We can similarly calculate the connected components of $\Maps(NX,SO(3))$, which is again given by the homotopy group $\pi_3(SO(3))\cong\IZ$. In particular, this space has infinitely many components.

On the other hand, recall that the space of ACM3S is the total space of a bundle, \eqref{eq:AC3Sfibn},
\begin{equation}
\scT\rightarrow \scC\rightarrow\scS\,,
\end{equation}
where the fibre is the space of trivialisations, $\scT=\Maps(NX,SO(3))$ and the base is the space of splittings $\scS= \Gamma(NX,\cG_2(\varphi))$. We want to know whether this bundle is trivial or not.  If we assume that it is a trivial bundle, \[\Gamma(NX,\cV_2(TNX))\cong \Maps(NX,SO(3))\times \Gamma(NX,\cG_2(\varphi)) \;,\] then we would conclude that 
\[ \pi_0(\Gamma(NX,\cV_2(TNX)))\cong \pi_0(\Maps(NX,SO(3)))\times\pi_0(NX,\cG_2(\varphi)) \; .\] This can not be, however, since the left-hand side is a finite set, and the right-hand side is infinite. Therefore, even in this very simple system the topological structure of the space of ACM3S's is non-trivial.

\section{Conclusions and outlook}\label{sec:conc}

In this paper we reviewed established results guaranteeing that $G_2$ structure manifolds admit a further reduction of structure group, first to $SU(3)$, then to $SU(2)$.  This reduction can be thought of as being topological in that geometric features of the $G_2$ structure, parallel transport for instance, need not respect this reduction. We saw, unsurprisingly, that compatibility of the geometry with this reduction leads to supersymmetry enhancement of the physics and, in the case of heterotic supergravity we explicitly analysed these compatibility conditions for an $SU(3)$ reduction.

For $SU(3)$ reductions, there is an induced foliation of the underlying seven manifold and by decomposing the field equations into longitudinal and transverse components with respect to this foliation, we found that the longitudinal components of the fields can be seen as measuring the flow along the foliation \emph{up to gauge transformations}. In other words, looking at the geometric features of the reduction of structure group allows us to interpret the seven-dimensional field equations as non-trivial flows of six-dimensional geometries.  Further, this analysis allowed us to precisely identify the conditions for supersymmetry enhancement, from $N=1$ to $N=2$, in terms of the $SU(3)$ structure, in complete agreement with the work in \cite{Gran:2005wf,Gran:2007kh,Gran:2016zxk}, and we used this to observe that an example from the literature \cite{Fernandez:2008wla}, had an hitherto unrecognised enhancement of supersymmetry.

This relation between almost contact structures and flows of six-dimensional geometries may have interesting consequences for the deformation theory of both six- and seven-dimensional geometries. From the six-dimensional perspective, one can consider flows in which the $SU(3)$ structure fails to satisfy the relevant equations of motion (for instance, Strominger-Hull in the heterotic case), but where this failure is cancelled by the change along the flow, leading to well-behaved seven-dimensional solutions. These kind of constructions have been considered in the context of domain walls, \cite{delaOssa:2014lma}, for instance, but the generality of the ACS may be leveraged to consider more complicated scenarios involving subtle interplay between six- and seven-dimensional geometries, such as intersecting networks of domain walls.  

Returning to the purely seven-dimensional setting, ACSs may be of use in the study of deformations of a $G_2$ structure manifold. In this case, the reduction of structure group can be thought of as an extra redundancy in the system, in other words a symmetry, and deformations can be expected to come in representations of this symmetry. Such considerations have often been used in physics. {In the case at hand,} we expect this will have non-trivial consequences for the difficult study of finite deformations of compactifications on $G_2$ structure manifolds.  In fact, looking further ahead, it may be possible to use these structures to study the infinite distance in moduli space, through decompactifying the foliation direction, for instance. That is, if we suppose that the foliation has {compact} leaves\footnote{In general, the leaves of a foliation need not be compact, so this procedure will not be sensible for every ACS.}, then we can consider the limit in the geometry where these leaves go to infinite length.{ This is reminiscent of Donaldson's study of adiabatic limits of Kovalev--Lefschetz fibrations \cite{2016arXiv160308391D}, and it would be interesting to make this connection more precise. From a physicist's perspective, the } usual stringy considerations \cite{Ooguri:2006in} imply that we ought to find a massive tower of states becoming light in this limit, and these states ought to correspond to excitations along the foliation. {What this tower of states consists of depends on which theory we study.} In this paper, we studied the decomposition of the fields of heterotic supergravity into longitudinal and transverse components, as well as the torsion of the $SU(3)$ structure connection, and we expect these results will be necessary in pursuing this idea {in the heterotic context. Moreover, it would be interesting to perform similar studies of, for example, M-theory compactified on manifolds with $G_2$ holonomy.}

A feature of importance for the future prospects of this research is that almost contact structures are abundant, indeed there is automatically an infinite dimensional family of them. It is, however, unclear whether each of these should truly be regarded as ``different'' from the physics perspective. It may perhaps be more fruitful to consider the further reduction of structure group to $SU(2)$ for these questions. We saw in this paper that this further reduction is accomplished by an $SO(3)$ triple of vector fields. This is the generic minimum structure group for a $G_2$ structure manifold, since any further reduction implies the manifold is, in fact, parallelizable i.e. that the structure group is trivial. 

Although the existence of an almost contact metric three-structure, an ACM3S, (and hence reduction of structure group to $SU(2)$) has been known since the middle of the last century, it seems the study of the space of such reductions has not been undertaken thus far.  In this paper we initiated the study of this space, which we identified as a space of sections on a bundle naturally associated to the principal $G_2$ frame bundle. It is, consequentially, infinite dimensional and has non-trivial topology in its own right, including being a locally trivial fiber bundle. We saw in examples that this may be non-trivially fibred and gave a rough argument for why this might be expected to hold in general. Further detailed analysis of this space, {for example for compact $G_2$ manifolds}, may prove to be interesting.

From the mathematical perspective, the space of these ACM3Ss, which we have denoted $\mathscr{C}$, depends only on the isomorphism class of the $G_2$ frame bundle. In this sense, it is a rather coarse $G_2$ invariant. An immediate question to ask is whether it truly depends on this bundle, or if it, in fact, depends only on the unreduced tangent bundle of the manifold. Perhaps of greater interest would be to find a refinement of this space, which somehow encodes geometric features of the $G_2$ structure, including the metric and covariant derivative, for instance.

Some physics considerations may point the way to constructing such a space.  Indeed, what is lacking in our analysis is any consideration of whether there is any ACM3S that is preferred. It is natural, from the physics perspective, that not all structures are created equal, so an important open problem is to discover a principle to differentiate between them. {As we have alluded to, there is a connection between ACM(3)S and supersymmetry, that one might hope to build on to discover such a principle. Let us make some arguments in favour of this.}

One line of reasoning is to relate the triplet of vectors to spontaneously broken supersymmetry, comparable to the approach in \cite{KashaniPoor:2013en}. This is a natural connection to make, because the vector fields induce spinors via clifford multiplication on the canonical $G_2$ spinor.  One might hope that these correspond to massless spinors, in the effective theory of a compactification scenario for instance, but  {unless supersymmetry is enhanced} they will correspond to particles with a mass measured by the $G_2$ Dirac operator. This suggests that this mass is the quantity to minimise {to find the preferred ACM3S, in the general case}.

In fact, this proposal has some ambiguities. There is no reason to expect a generic nowhere vanishing vector to induce an eigenspinor. Consequentially, a given AC(3)S will induce a linear combination of massive fields that can not be teased apart using the tools at hand. It may be too naive, then, to think of the triplet in an ACM3S as corresponding to particles of the effective theory.  Nevertheless, if we view that ACM3S as part of the data in defining an enhanced supersymmetric vacuum, then variations of ACM3S would correspond to vectors in the space of field configurations. Considering the variations of e.g. a superpotential would lead to Euler-Lagrange equations and it is these solutions that one might expect to single out preferred structures.  In fact, it might well be expected that the critical locus would be an interesting invariant of the $G_2$ geometry and thus be of independent interest to mathematicians.

Indeed, we discovered that integrability of the bundles appearing in the construction of the ACM3S is related to interesting aspects of the $G_2$ structure: associative and coassociative cycles. These have been much studied in the context of $G_2$ holonomy manifolds, impetus in the physics community coming from BPS states in M-theory compactifications, for instance.  When the $G_2$ structure is not, in fact, closed and coclosed, then identifying the relevant BPS states is a still open question. There is some hope that the ACM3S studied here may help shed light on this, once one has identified the correct action principle.

The relation between ACM3Ss and associative cycles was explored further in examples. In particular, we looked at the local neighbourhood of an associative three-cycle in a $G_2$-holonomy manifold. It was here that we were able to explicitly show that the space of ACM3Ss, $\mathscr{C}$, was non-trivially fibred over the space of trivial, associative rank 3 subbundles of the tangent bundle. We also saw that we could always find an ACM3S that was tangent to the associative three-cycle, but that if we fixed the behaviour at the boundary of this manifold, then there are possible obstructions to doing so. Re-inserting this local picture into a compact $G_2$-holonomy manifold is therefore non-trivial. It would be very interesting to invert this problem and consider how the global topology of a compact manifold affects the possibility of an ACM3S lying tangent to a given associative cycle (or, dually, for the transverse bundle to lie tangent to a coassociative).

One way to approach this question is, of course, to study more examples. There are several classes to explore: further nilpotent examples from the list of \cite{delBarco:2020ddt},  $T^3$ bundles over $K3$  \cite{Fernandez:2008wla}, \cite{Clarke:2020erl}, and the very recently constructed examples of heterotic $G_2$ systems on contact Calabi--Yau seven-manifolds \cite{Lotay:2021eog}. Of equal import are the different constructions of compact $G_2$ holonomy manifolds: Joyce orbifolds \cite{joyce1996:1,joyce1996:2} and (extra) twisted connected sums \cite{kovalev20,Corti:2012kd,Corti:2013,2018arXiv180909083N}. Here, we have initiated studies of ACM3S on Joyce orbifolds, and shown that they come equipped with a canonical AC(3)S. We expect similar results to hold for TCS manifolds. Non-compact $G_2$ manifolds, such as Bryant and Salamon's seminal examples \cite{bryant89} and the more recent constructions \cite{Foscolo:2017vzf,Acharya:2020vmg}, also have clear links with $SU(3)$ structures and ACS. Determining the space $\scC$ on some of these example geometries would be interesting.

One final possibility that we would like to highlight is the utility of these structures in {applications of localisation techniques in quantum field theories (see \cite{Pestun:2016zxk} for a review)}. Indeed, localisation techniques have been successfully applied to classes of 7-manifold admitting particular kinds of contact structure, \cite{Minahan:2015jta,Polydorou:2017jha,Rocen:2018xwo,Iakovidis:2020znp}. Extending these results to a general $G_2$ structure manifold would be incredibly interesting, but it thus far remains intractable. It is plausible that the {almost contact} structures considered here will of be use, in that they are similar (but weaker) to the {contact structures already used in Refs.~\cite{Minahan:2015jta,Polydorou:2017jha,Rocen:2018xwo,Iakovidis:2020znp}}, while being ubiquitous.  

In short, we have found that these almost contact (three) structures have tantalising connections to many active research in both physics and mathematics. We think that the tools and results we reviewed and established here will be necessary in fleshing out these surprising relations.

\section*{Acknowledgements}
The authors would like to thank Eirik Svanes for initial collaboration on this project. 
We would also like to thank  Marc-Antoine Fiset, Mateo Galdeano Solans and Maxim Zabzine for discussions.
The research of ML and MM is financed by Vetenskapsr\aa det under grant number 2016-03873, 2016-03503, and 2020-03230.
\newpage


\appendix

\section{$SU(3)$ structures}\label{app:su3struct}

Let $X$ be a 6 dimensional Riemannian manifold with metric $g$.  An $SU(3)$ structure on $X$ is defined by a triple $(X,\omega,\Omega)$, where $\omega$ is a positive non-degenerate  globally well defined real two form, and $\Omega$ is a locally decomposable nowhere vanishing globally well defined decomposable three form.  The forms $\Omega$ and $\omega$ satisfy
\be
\omega\wedge\Omega = 0~.
\label{eq:comp}
\ee
The real part of the form $\Omega$ determines an almost complex structure $J$ on $X$. 
With respect to $J$, $\Omega$ is a $(3,0)$-form and, by equation \eqref{eq:comp}, $\omega$ is a $(1,1)$ form.  In fact $\omega$ is a hermitian form on $X$. 
The almost complex structure together with $\omega$  determine a hermitian metric on $X$
\[
g_{mn} = \omega_{mp}\, J^p{}_n = - J^p{}_m\, \omega_{pn}
~.\]
There is a unique, up to a constant, invariant volume form
which can be written as
\be
\dd{\rm vol} = \frac{1}{3!}\, \omega\wedge\omega\wedge\omega = \frac{i}{||\Omega||^2}\, \Omega\wedge\overline\Omega~,
\qquad
||\Omega||^2 = \overline\Omega\lrcorner\Omega
~.\label{eq:su3compBis}
\ee
As we will see later, the $SU(3)$ structure which is natural to a manifold with a $G_{2}$ structure satisfies $||\Omega||^2 = 8$.  This means that the scale invariance of this $SU(3)$ is reduced to an invariance under phase changes of $\Omega$.

The exterior derivative of the form $\omega$ and $\Omega$ can be decomposed into representations of $SU(3)$ as follows
\begin{align}
\dd\omega &= - \frac{12}{||\Omega||^2}\, {\rm Im}(W_0\, \overline\Omega) + W_1\wedge\omega + W_3~,
\label{eq:su3om}
\\
\dd\Omega &= W_0\, \omega\wedge\omega + W_2\wedge\omega + \overline \theta\wedge\Omega~,
\label{eq:su3Om}
\end{align}
where the forms $\{W_0, \theta, W_1, W_2, W_3\} $ 
are the torsion classes. $W_0$ is a complex function, $W_1$ is a real one form, $\theta$ a $(1,0)$ form,  
$W_2$ is a complex primitive $(1,1)$ form, and $W_3$ is a real primitive three form type $(2,1)+(1,2)$.


\section {$SU(2)$ structures}\label{sec:su2}

Let $S$ be a four dimensional manifold with metric $g$.
An $SU(2)$ structure on $S$ is defined by a triple $(S,\omega,\Omega)$, where $\omega$ is a positive, non-degenerate,  globally well defined real two form, and $\Omega$ is a locally decomposable, nowhere vanishing, globally well defined two form.  The forms $\Omega$ and $\omega$ satisfy
\be
\omega\wedge\Omega = 0~.
\label{eq:compsu2}
\ee

There is a unique, up to a constant, invariant volume form
\be
\dd{\rm vol}_S = \frac{1}{2}\, \omega\wedge\omega 
= \frac{1}{||\Omega||^2}\, \Omega\wedge\overline\Omega~,
\qquad
||\Omega||^2 = \overline\Omega\lrcorner\Omega
~.\label{eq:su2comp}
\ee
While keeping the scaling of $\Omega$ as an invariance of the theory, in the $G_2$ setting we are interested in, we will have $||\Omega||^2 = 1$.

We will not, in this paper, need the torsion classes of the $SU(2)$ structure, which is found by decomposing the  exterior derivative of the form $\omega$ and $\Omega$ into representations of $SU(2)$. The reader is referred to \cite{Behrndt:2005im}, and also \cite{Bovy:2005qq,ReidEdwards:2008rd,Louis:2009dq,KashaniPoor:2013en} for more details on this point.

Just as in the $SU(3)$ structure discussed in the previous subsection, we have that the complex two-form defines an almost complex structure on $TX$. In fact there are several almost complex structures on an $SU(2)$ structure manifold, as we now explain. We may always, as we do in the main discussion of this paper, trade the complex two-form $\Omega$ for two real two-forms. Let
\be
\omega^1 = {\rm Re} \Omega \; , \; \omega^2 = {\rm Im} \Omega  \; , \;  \omega^3 = \omega\; .
\ee
Then $\{\omega^1, \omega^2, \omega^3\}$ transform as an $SU(2)$ triplet, and any complex combination of these forms defines an almost complex structure. Moreover, we may associate an almost complex structure to each $\omega^{\alpha}$:
\be
(J^{({\alpha})})^a{}_b = g^{ac} \omega^{\alpha}_{bc} \; .
\ee
It can be shown that $J^{({\alpha})} J^{(\beta)} = - \delta^{\alpha \beta}\bf{1}$ follows from $2\,\omega^{\alpha} \w \omega^{\beta} = \delta^{\alpha\beta} \dd{\rm vol}_S $, which in turn follows from \eqref{eq:su2comp}. The almost complex structure $J^{(\alpha)}$ is integrable if the corresponding two-form is closed. In the case that all $\omega^{\alpha}$ are closed, the four-dimensional manifold is hyper-K\"ahler, and the $SU(2)$ transformations that rotate the $\omega^{\alpha}$ correspond to hyper-K\"ahler rotations of the complex structure. If the structure group is proper $SU(2)$, and not a subgroup thereof, the closure of all $\omega^{\alpha}$ implies that $X$ is a K3 surface.


\bibliographystyle{JHEP}

\bibliography{bibliography}

\end{document}